\documentclass[acmlarge, authorversion]{acmart}

\makeatletter
\newcommand{\confshort}{\acmConference@shortname}
\newcommand{\conffull}{\acmConference@name}
\newcommand{\confdate}{\acmConference@date}
\newcommand{\confloc}{\acmConference@venue}
\AtBeginDocument{
  \fancypagestyle{firstpagestyle}{
    \fancyhead{}%
    \fancyfoot[C]{}%
  }
  \fancyhf{}
  \fancyhead[LO]{\@headfootfont\shorttitle}%
  \fancyhead[RE]{\@headfootfont\@shortauthors}%
  \fancyhead[LE]{\@headfootfont\footnotesize \confshort, \confdate, \confloc}%
  \fancyhead[RO]{\@headfootfont\footnotesize \confshort, \confdate, \confloc}%
  \fancyfoot[C]{}%
}
\makeatother
\acmBooktitle{\conffull\@ (\confshort), \confdate, \confloc}

\AtBeginDocument{%
  }

\copyrightyear{2026}
\acmYear{2026}
\setcopyright{cc}
\setcctype{by-nc-nd}
\acmConference[FAccT '26]{The 2026 ACM Conference on Fairness, Accountability, and Transparency}{June 25--28, 2026}{Montreal, QC, Canada}
\acmBooktitle{The 2026 ACM Conference on Fairness, Accountability, and Transparency (FAccT '26), June 25--28, 2026, Montreal, QC, Canada}
\acmDOI{10.1145/3805689.3812347}
\acmISBN{979-8-4007-2596-8/2026/06}

\usepackage{longtable}
\usepackage{array}

\usepackage{color}

\usepackage{float}
\usepackage{multirow}
\usepackage{tablefootnote}
\usepackage{tikz}
\usetikzlibrary{shapes, backgrounds}
\usepackage{enumitem}
\usepackage{listings}
\usepackage{setspace}
\usepackage{caption}

\begin{document}


\title[Why AI Harms Can't Be Fixed One Identity at a Time]{Why AI Harms Can't Be Fixed One Identity at a Time: \\What 5300 Incident Reports Reveal About Intersectionality}


\author{Edyta Paulina Bogucka}
\orcid{0000-0002-8774-2386}
\affiliation{
  \institution{Nokia Bell Labs}
  \city{Cambridge}
  \country{United Kingdom}}
\email{edyta.bogucka@nokia-bell-labs.com}

\author{Sanja \v{S}\'{c}epanovi\'{c}}
\orcid{0000-0002-1534-8128}
\affiliation{
  \institution{Nokia Bell Labs}
  \city{Cambridge}
  \country{United Kingdom}}
\affiliation{
  \institution{University of Oxford}
  \city{Oxford}
  \country{United Kingdom}}
\email{sanja.scepanovic@nokia-bell-labs.com}

\author{Daniele Quercia}
\orcid{0000-0001-9461-5804}
\affiliation{
  \institution{Nokia Bell Labs}
  \city{Cambridge}
  \country{United Kingdom}}
\affiliation{
  \institution{Politecnico di Torino}
  \city{Turin}
  \country{Italy}}
\email{quercia@cantab.net}

\renewcommand{\shortauthors}{Bogucka et al.}

\begin{abstract}
    AI risk assessment is the primary tool for identifying harms caused by AI systems. These include intersectional harms, which arise from the interaction between identity categories (e.g., class and skin tone) and which do not occur, or occur differently, when those categories are considered separately. Yet existing AI risk assessments are still built around isolated identity categories, and when intersections are considered, they focus almost exclusively on race and gender. Drawing on a large-scale analysis of documented AI incidents, we show that AI harms do not occur one identity category at a time. Using a structured rubric applied with a Large Language Model (LLM), we analyze 5,300 reports from 1,200 documented incidents in the AI Incident Database, the most curated source of incident data. From these reports, we identify 1,513 harmed subjects and their associated identity categories, achieving 98\% accuracy. At the level of individual categories, we find that age and political identity appear in documented AI harms at rates comparable to race and gender. At the level of intersecting categories, harm is amplified up to three times at specific intersections: adolescent girls, lower-class people of color, and upper-class political elites. We argue that intersectionality should be a core component of AI risk assessment to more accurately capture how harms are produced and distributed across social groups.
\end{abstract}



\received{13 January 2026}
\received[revised]{25 February 2026}
\received[accepted]{15 April 2026}

\maketitle

\section{Introduction}
\label{sec:introduction}

In 2018, the Argentinian province of Salta approved the deployment of a machine-learning system to predict adolescent pregnancy. Trained on health, education, and housing data, the system generated individual risk scores. Its deployment triggered public backlash after news reports showed that the resulting harms were not evenly distributed \cite{AIID_Incident188_2018}. Girls were singled out for risk scoring and follow-up, while boys were excluded entirely, placing responsibility for pregnancy prevention on girls. Girls in rural and Indigenous communities were more likely to receive in-person follow-ups such as household visits than girls in urban areas. Pre-adolescent girls were labeled years in advance, even though they had not yet engaged in adolescent behavior.

This unequal distribution of harms is commonly explained using the theory of intersectionality. The concept was originally introduced by Crenshaw to describe how legal and policy frameworks failed to account for the combined effects of race and gender \cite{Crenshaw1991}. Subsequent scholarship emphasizes that intersectionality describes not only interactions among identity categories, but also the forms of social power that structure inequality such as racism, sexism, and classism \cite{HillCollins2002, Nash2019, Goodwill_2021}. The Salta case illustrates this logic in practice. The system caused harm not because of age, gender, ethnicity, or geography on their own, but because of how these factors played out within powerful institutions: public health bodies, welfare agencies, and regional governments. This dynamic is captured in a widely cited definition of intersectionality by \citet{BrahPhoenix2004_IntersectionalityDefinition}, who describe it as ``the complex, irreducible, varied, and variable effects which ensue when multiple axes of differentiation~\textemdash~economic, political, cultural, psychic, subjective and experiential~\textemdash~intersect in historically specific contexts''. Throughout this paper, we refer to these axes of differentiation as \emph{identity categories} (such as race, gender, or age), and to their specific instantiations as \emph{identity values} (such as Black, woman, or adolescent).

Despite broad recognition of intersectionality in theory, empirical research on AI-related harms has applied it narrowly. Prior work has largely focused on a small number of isolated identity categories \cite{Bauer_2021}, most commonly race and gender \cite{IntersectionalHCI_2017, AIMargins_2022, DiversityInclusionIncidents_2025}. While this literature has established that intersectional harms exist \cite{UnfairTreatment_2022}, it offers limited insight into how they are distributed across many identity categories and AI uses. As a result, the empirical bases used in AI risk assessment such as fairness benchmarks \cite{GermanCreditData_1994, AdultDataSet_1996, COMPASS_2016} and incident taxonomies \cite{harmTaxonomy_2024} may overlook identity configurations that are common in AI incidents. To make progress on this gap, we ask:

\begin{itemize}
    \item[(RQ\textsubscript{1})] Which identity categories and their corresponding values appear most often in AI incidents reported in the news, and what kinds of harm are linked to them?
    \item[(RQ\textsubscript{2})] How do intersecting identity categories affect the likelihood and types of harm caused by AI systems?
\end{itemize}

\noindent To address these questions, we make three main contributions:

\begin{enumerate}[leftmargin=*]
    \item \textbf{We develop a systematic approach for identifying and characterizing intersectional AI harms in incident reports (\S\ref{sec:methodology}).} 
    The approach specifies a structured rubric for identifying harmed subjects, relevant identity categories and values, and associated harm types. It can be applied either manually or with the support of Large Language Models (LLMs) to analyze heterogeneous and unstructured reports at scale.
    
    \item \textbf{We apply this approach to a large set of documented AI incidents to produce an empirical characterization of intersectional AI harms (\S\ref{sec:results}).} 
    We analyze 5,300 incident reports across 1,200 unique incidents from the AI Incident Database \cite{AIID}, identifying 1,513 harmed subjects with an accuracy of 98\%. Across single identity categories implicated in these harms, age (32\%) and political identity (27\%) appear as frequently as race (25\%) and gender (24\%). However, we show that AI harms do not occur one identity category at a time. The most common intersectional harms involve nationality and political identity through deepfake targeting (12\%), age and gender through sexualized profiling (10\%), and nationality and class through denial of services (6\%). At these intersections, people identified as upper-class political elites, adolescent girls, and lower-class people of color appear up to three times more often than would be expected if each identity category contributed to harm on its own. 
    
    \item \textbf{We identify recurring patterns and the institutional contexts through which intersectional AI harms arise (\S\ref{sec:results}).}
    Through thematic analysis, we find six such patterns (six specific ways AI systems act on identity attributes): sexualizing, steering, matching, inferring, gating, and manipulating (\S\ref{subsec:results_rq1}). These interact with institutions of social power to produce intersectional harms: algorithmic suspicion, sexualized exploitation, environmental exposure, political manipulation, and militarized state violence (\S\ref{subsec:results_rq2}).   
\end{enumerate}

Building on these contributions, we discuss how intersectional AI harms emerge across both high-visibility and everyday systems, consider the implications for AI risk assessment, and discuss the limitation of using news as sources of our incidents (\S\ref{sec:discussion}). To support researchers in advancing this research direction, we have publicly released our approach at \textbf{\url{https://social-dynamics.net/ai-risks/intersectional}}.
\section{Related Work}

Next, we review prior work on AI-related harms through two complementary lenses: how harms manifest across intersecting identities (\S\ref{subsec:identity_intersections}), and how they vary across AI use cases (\S\ref{subsec:use_cases}).

\subsection{Studying AI Harms Across Identity Intersections}
\label{subsec:identity_intersections}

Researchers in human-computer interaction (HCI) and AI ethics have increasingly drawn on intersectionality theory \cite{Crenshaw1991, HillCollins2002, Nash2019, Goodwill_2021} to examine how AI systems distribute harms unevenly across marginalized and privileged groups and reinforce forms of social power such as racism, sexism, and classism \cite{IntersectionalHCI_2017, UnfairTreatment_2022}. In practice, however, this work has largely focused on intersections of only two identity categories \cite{Bauer_2021}, most often race and gender \cite{IntersectionalHCI_2017, AIMargins_2022, DiversityInclusionIncidents_2025}. For example, \citet{GenderShades_2018} showed that commercial facial analysis systems misclassified darker-skinned women at substantially higher rates than lighter-skinned men. Similarly, audits of CV screening systems found that LLMs favored CVs with white-sounding names 85\% of the time, while never favoring Black male names over white male names \cite{ResumeScreening_2025}. These findings illustrate how harms at specific intersections such as allocative harms affecting Black men in hiring, remain invisible when race or gender is analyzed in isolation.

Across 526 papers published between 2018 and 2021 at FAccT and AIES, race and gender each appear in nearly 20\% of papers, while categories such as disability, socioeconomic status, and religion appear in fewer than 3\% \cite{AIMargins_2022}. Analyses of widely used fairness datasets further show that attributes related to disability, religion, health and belief are largely absent, and that socioeconomic variables, while more available, are rarely used empirically \cite{LazyDataPractices_2024}. Recognizing this narrow focus, prior work has called for greater attention to underexamined intersections, such as class and race \cite{IntersectionalHCI_2017} or disability and age \cite{UnfairTreatment_2022}, though addressing them remains challenging in practice \cite{IntersectionalSIGCHI_2018}. For example, the Political Deepfakes Incident Database documents harms at intersections of class and political identity using manually collected cases involving public figures. \cite{politicalDeepfakes2024}. Such harms range from discrimination to targeted violence to financial loss.

Recent work has begun to overcome constraints on studying intersections involving more than two identity categories by leveraging LLMs. \citet{DisabilityHiring_2025} generated hiring conversations for school teacher and software engineer positions across six LLMs, constructing candidate profiles that explicitly combine disability, gender, and caste. They found that adding disability alone increased harmful conversation content by up to 58 times relative to baseline profiles, while further adding gender and caste increased harm by an additional 10 to 51\%. They also showed that candidates at these combined intersections were exposed to new types of harm: being framed as inspirational because of their identity rather than their qualifications (``inspiration porn''), portrayed as possessing exaggerated abilities (``superhumanization''), or valued primarily for fulfilling workplace diversity goals rather than for their professional merit (``tokenism'').

\subsection{Studying AI Harms Across Use Cases}
\label{subsec:use_cases}

Much of the empirical literature on AI harms focuses on use cases that would be classified as high risk under the EU AI Act, that is, systems deployed in sensitive domains where failure or misuse can cause significant harm \cite{EUAIAct2024}. Examples include hiring \cite{ResumeScreening_2025}, criminal justice \cite{HillCollins2002}, benefit allocation \cite{Redden02092020, WiredSuspicionMachine2023, LighthouseSuspicionMachineMethodology2023} and targeted advertising \cite{TargetedAdvertising_2023}. While this focus has yielded rich analyses, it has concentrated research attention on a narrow set of benchmark datasets and tasks. For example, much fairness research relies on COMPAS dataset for recidivism assessment \cite{COMPASS_2016}, German Credit for credit risk assessment \cite{GermanCreditData_1994}, and UCI Adult for income prediction \cite{AdultDataSet_1996}. This reliance on a few benchmarks limits coverage of the diversity of AI deployments and the harms they produce \cite{AIMargins_2022, LazyDataPractices_2024}.

A growing body of qualitative work has examined harms in seemingly low risk, everyday algorithmic systems, where users encounter AI quietly and repeatedly in routine interactions. \citet{UnfairTreatment_2022} documented nearly 100 firsthand accounts of unfair treatment across everyday systems such as shopping platforms, with harms reported along axes of gender, sexual orientation, race and ethnicity, disability, and geographic location. Across these identities, participants most frequently described being forced into inaccurate or adjacent categories by automated systems (``enforced categorization''), or being rendered unrecognizable altogether (``symbolic annihilation'' \cite{SymbolicAnnihilation_2021}). \citet{BlackWriting_2025} examined AI-supported writing technologies and showed that even mundane tools such as grammar checkers and autocorrect systems can generate identity-related harms. They documented these harms along race and language: Black users reported that these tools failed to recognize African American Vernacular English, flagged culturally common expressions as errors, and framed their linguistic practices as unprofessional, leading to experiences of exclusion and cultural erasure.

In parallel with qualitative research, recent quantitative, data-driven work has examined AI harms across use cases at scale using large collections of incidents \cite{riskAtlas2024, DecodingIncidents_2024, DiversityInclusionIncidents_2025}. These studies draw on volunteer-submitted incident reports curated in public databases, reflecting the limited availability of structured top-down incident reporting mechanisms such as those envisioned under the EU AI Act \cite{EUAIAct2024}. For example, \citet{DiversityInclusionIncidents_2025} analyzed incident reports from the AI Incident Database \cite{AIID} and the AI, Algorithmic, and Automation Incidents and Controversies database \cite{AIAAIC} collected by mid-2023. These studies identified 18 diversity attributes implicated in AI incidents, including facial features, culture, and accent. However, across both databases, reported identity-related harms most commonly involved bias or discrimination along the same identity categories that recur across much of the literature: race and gender.

\smallskip
\noindent\textbf{Research gap.} Although intersectional AI harms are well established in theory, existing approaches to AI risk assessment focus on a narrow subset of isolated identity categories; when intersections are considered, they are almost exclusively limited to race and gender. We address this limitation by analyzing 5,300 AI incident reports to examine identity-related harms across a broader set of identity categories and less-explored intersections.

\section{Methods for Analyzing Identity-Related AI Harms}
\label{sec:methodology}

\begin{figure}[t!]
    \centering
    \includegraphics[width=\linewidth]{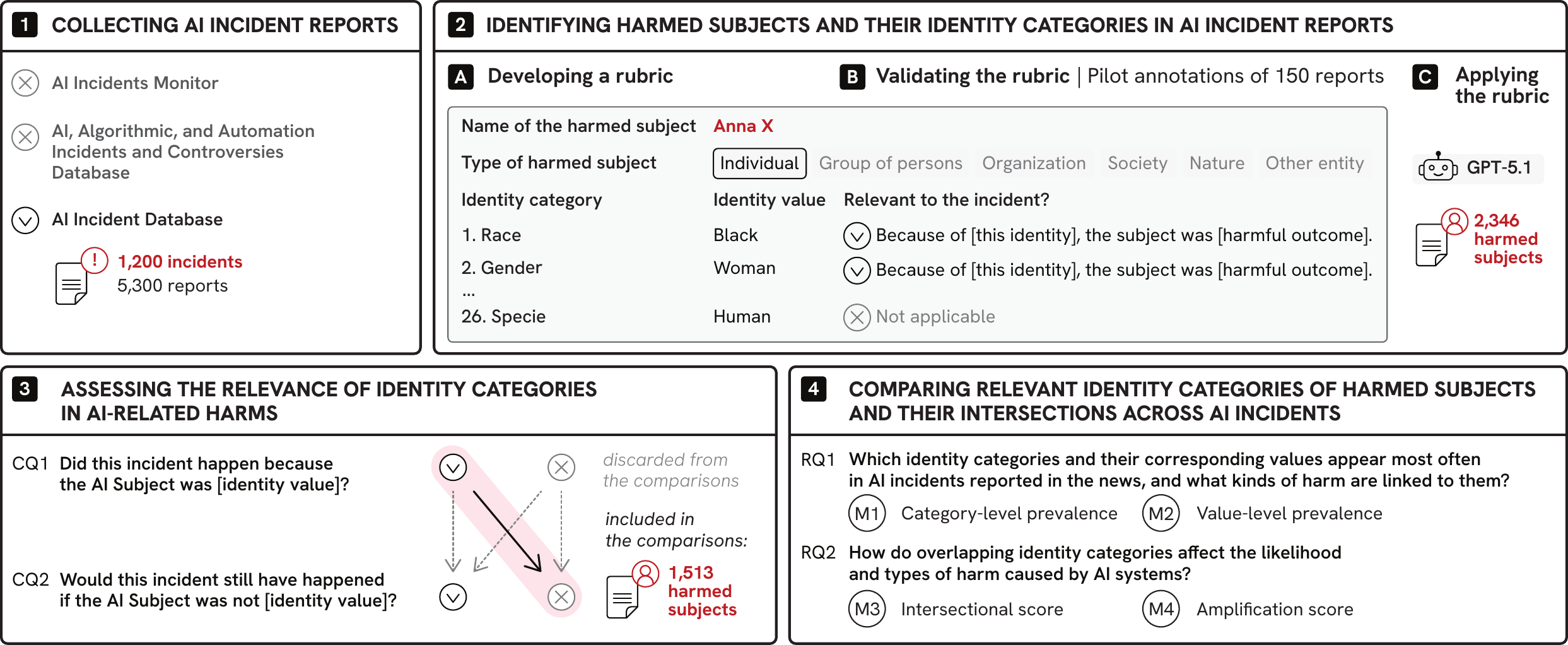}
    \caption{\textbf{Overview of our four-step methodology for identifying and analyzing intersectional AI harms in AI incidents.} The approach combines large-scale incident collection (1), a rubric for identifying the identities of harmed subjects (2A–B), LLM-assisted extraction of these identities using the rubric (2C), and counterfactual relevance assessment (3). Together, these steps enable the systematic identification of harmed subjects, their identity categories and intersections, and the concrete harms attributable to identity-specific factors across the diverse AI uses covered in the incident reports (4).}
    \Description{A four-part flow diagram showing the methodology for identifying and analyzing intersectional AI harms across reported incidents.

    Step 1, ``Collecting AI Incident Reports'', shows data sources feeding into an AI Incident Database, including the AI Incidents Monitor and the AI, Algorithmic, and Automation Incidents and Controversies Database, totaling 5,300 reports and 1,201 incidents.
    
    Step 2, ``Identifying Harmed Subjects and Their Identity Categories'', is subdivided into three stages: developing a rubric, validating the rubric with pilot annotations of 150 reports, and applying the rubric using an LLM (GPT-5.1). A form-like example illustrates how a harmed subject (``Anna X'') is annotated with subject type (individual, group, organization, society, nature, or other) and multiple identity categories such as race, gender, and specie. Each identity value (for example, Black, woman, human) is marked as either relevant or not relevant to the harm using a counterfactual statement (``Because of this identity, the subject was harmed''). This step identifies 2,346 harmed subjects, some of which are later discarded.
    
    Step 3, ``Assessing the Relevance of Identity Categories'', presents two counterfactual questions: whether the incident occurred because the subject had a given identity value, and whether it would still have occurred without that identity. Based on the answers, harmed subjects are either included in or excluded from further comparison, resulting in 1,513 retained subjects.
    
    Step 4, ``Comparing Relevant Identity Categories and Their Intersections'', summarizes the analysis outcomes. It defines two research questions: which identity categories and values appear most often in AI harm cases and what harms are linked to them, and how overlapping identity categories affect the likelihood and types of harm. Corresponding metrics are listed: category-level prevalence, value-level prevalence, an intersectional score, and an amplification score.
    
    Overall, the figure depicts a linear pipeline from incident report collection, through structured identity annotation and relevance filtering, to quantitative analysis of identity-based and intersectional AI harms.}
    \label{fig:methodology}
\end{figure}

To analyze identity-related AI harms, we followed a four-step approach (Figure~\ref{fig:methodology}, Steps~1--4). We first collected AI incident reports (\S\ref{subsec:step1}), then identified harmed subjects and their identity categories within these reports (\S\ref{subsec:step2}). Because an incident may involve an identity category without being caused by it, we assessed category relevance using counterfactual evaluation (\S\ref{subsec:step3}), and finally compared relevant categories and their intersections across incidents (\S\ref{subsec:step4}).

\subsection{Collecting AI Incident Reports}
\label{subsec:step1}

Identity-related AI harms typically become visible not in controlled settings, but when deployed systems fail or harm populations. To identify harmed subjects and implicated identity categories, we rely on AI incident data, following \citet{DecodingIncidents_2024} and \citet{DiversityInclusionIncidents_2025}. We define three criteria for selecting data sources: (C1) preservation of original materials such as serious incident reports \cite{EUAIAct2024}, (C2) sufficient detail to support intersectional analysis, and (C3) collection and maintenance with human oversight. Appendix \ref{appendix:criteria} explains how these criteria were applied.

We evaluated three data sources commonly used in studies of AI harms \cite{riskAtlas2024, DecodingIncidents_2024, DiversityInclusionIncidents_2025}: the OECD AI Incidents Monitor (AIM \cite{OECD_AIM}; $K_{\text{incidents}} = 12,500$), the AI, Algorithmic, and Automation Incidents and Controversies database (AIAAIC \cite{AIAAIC}; $K_{\text{incidents}} = 2,100$), and the AI Incident Database (AIID \cite{AIID}; $K_{\text{incidents}} = 1,200$). We selected the AIID as the only source satisfying all three criteria.

The AIID is a curated repository of AI incident reports submitted by community members, researchers, and industry practitioners. Submissions are collected through a public web form where contributors provide incident metadata and the text of underlying report(s). Reports must originate from public online sources such as news media, court records, or company disclosures, though, in practice, most derive from news coverage. After submission, AIID editors review the materials and determine whether to include the incident in the database. 

This reporting pipeline introduces four main biases: (1) concentration in media-visible domains (e.g., communication, transportation) \cite{Richards2025}; (2) reliance on liberal or center-left news sources \cite{DecodingIncidents_2024}; (3) amplification of incidents involving public figures or politically identifiable groups through repeated media coverage \cite{politicalDeepfakes2024}; and (4) a focus on incidents occurring in the United States \cite{DecodingIncidents_2024}. We discuss these biases when interpreting our results in \S\ref{subsec:limitations}.

We then obtained a complete dataset from the AIID as of September 1, 2025. The dataset included incident identifiers, incident titles, descriptions, the number of associated reports, and the full text of each report, spanning 1,200 unique incidents and 5,300 reports. Incidents were documented by 1–58 reports ($\mu = 4$), with 61\% associated with more than 1 report. This structure provided sufficient detail to identify harmed subjects and extract identity categories while enabling comparison across incidents documented by multiple reports.

\subsection{Identifying Harmed Subjects and Their Identity Categories in AI Incident Reports}
\label{subsec:step2}

We proceeded to identify harmed subjects and their identity categories in each report. To do so, we developed a rubric, applied it at scale to the full dataset using an LLM, and validated the results.

\subsubsection{Developing a Rubric.}
\label{subsubsec:rubric}

We developed the rubric by reviewing foundational and subsequent research on intersectionality \cite{CombaheeRiverCollective1982, Crenshaw1991, HillCollins2002, Pauly_Wheel_1996, Goodwill_2021, Bauer_2021}, representation of users in HCI \cite{IntraCategoricalApproach_2005, IntersectionalHCI_2017, DisabilityHiring_2025}, and existing regulatory classifications relevant to AI harms \cite{EUAIAct2024, Hagendorff_2022}. The rubric consists of five components.

First, it identifies harmed subjects using their exact name or descriptor as it appears in the report text \cite{AIID}.

Second, the rubric defines a typology of harmed subjects (i.e., individuals, groups, organizations, societies, and nature), adapted from the EU AI Act \cite{EUAIAct2024}, a leading framework for AI governance.
 
Third, the rubric specifies a fixed set of 26 identity categories for each harmed subject and requires assigning a concrete value within a category when applicable. For example, the caste category can include values such as ``Brahmin'' and ``Dalit'' \cite{DisabilityHiring_2025}, and the education category can include values such as ``student'' or ``vocational trainee''. The category set was developed using an intercategorical approach \cite{IntraCategoricalApproach_2005}, treating identity categories as provisional analytical labels rather than fixed or exhaustive descriptions \cite{IntersectionalHCI_2017}. Categories were sourced through iterative review of seminal intersectionality and HCI literature \cite{CombaheeRiverCollective1982, Crenshaw1991, HillCollins2002, Pauly_Wheel_1996, Goodwill_2021, Bauer_2021, Neurodiversity_2025}. Where categories could take a wide range of values, the rubric specifies explicit grouping rules to ensure consistency in value assignment. For class, for example, the rubric adopts the definition from \citet{Ames2011} as ``a nexus of income level, educational attainment, and type of employment''. Since education is captured by its own category, the rubric requires grouping only income and occupation values into lower, middle, and upper class (e.g., ``gig worker'' to lower, ``small business owner'' to middle, and ``politician'' to upper class). The full list of categories, their exemplary values, grouping rules, and methodological process are provided in Appendix~\ref{appendix:categories_list}.

Fourth, the rubric outlines how to assess whether an identity category is relevant to an incident using two counterfactual questions: (CQ1) whether the incident occurred because the subject had this identity category, and (CQ2) whether the incident would likely have occurred if the subject were identical in all respects except this identity category.

Fifth, the rubric requires a description of the harm experienced by the subject due to their identity category.

Prior to its application at scale, we improved the rubric through pilot annotations. Two members of the research team independently applied preliminary versions of the rubric to a pilot set of 50 incident reports sampled to span different AI subject types and harm contexts. These annotations were compared and discussed in joint review sessions. Disagreements were mostly about relevance judgments for borderline cases where identity information was ambiguously related to harm. These were used to refine category definitions and improve the phrasing of the counterfactual relevance questions. This process was repeated iteratively until the rubric yielded stable annotations across pilot reports.

\subsubsection{Applying the Rubric.}

We then applied the rubric at scale to the incident reports by translating it into a structured prompt that can be executed by LLMs (provided in Appendix \ref{appendix:rubric_prompt}). To analyze our dataset, we used OpenAI's \textsc{GPT-5.1} via batch API processing. At the time of analysis, this model showed the strongest performance on capabilities relevant to our rubric, including instruction following (SWE-bench Verified benchmark), extraction of information from long documents (BrowseComp Long Context benchmark), and multi-step reasoning (GPQA Diamond benchmark). 

\begin{figure}[t!]
    \centering
    \includegraphics[width=\linewidth]{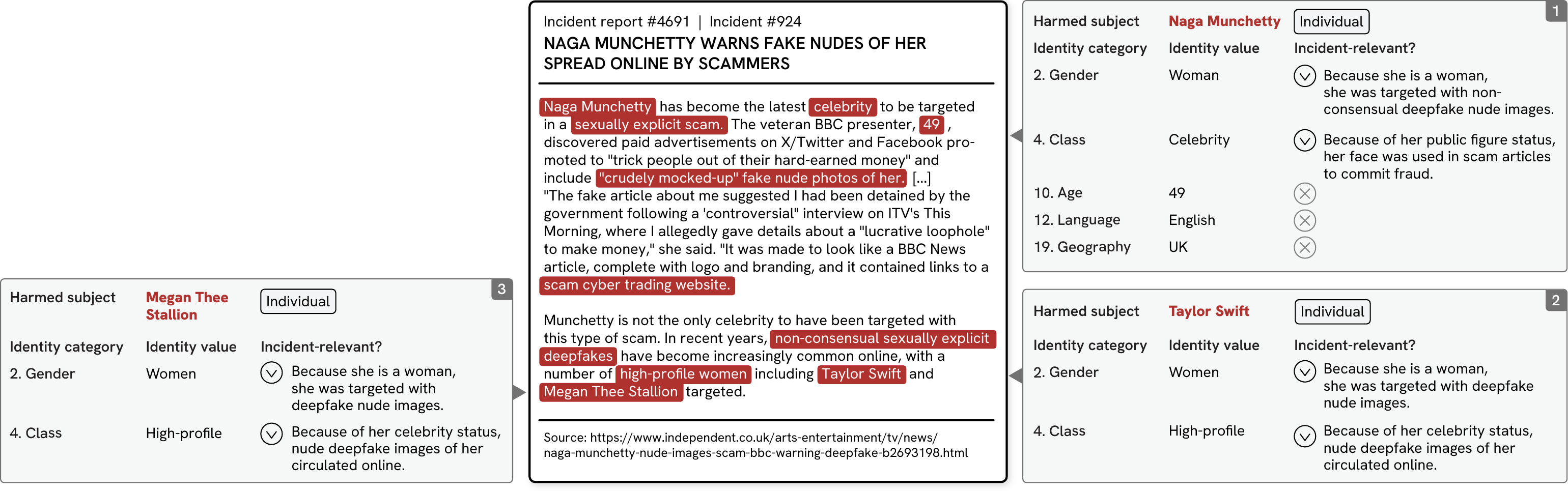}
    \caption{\textbf{Example of an incident report illustrating harmed subjects, their identity categories and values, and associated harm descriptions extracted by an LLM using our rubric.} The example of report 4691 highlights three harmed subjects whose identities combine female gender with upper-class roles such as media personalities and celebrities.}
    \Description{A figure showing an incident report (report 4691, incident 924) titled ``Naga Munchetty warns fake nudes of her spread online by scammers'', with key phrases highlighted in red. Alongside there are three structured annotation panels with identity categories and their values extracted by an LLM. The central panel displays the full report text, with highlighted
    entities including the names Naga Munchetty, Taylor Swift, and Megan Thee Stallion, and phrases such as ``sexually explicit scam'', ``non-consensual sexually explicit deepfakes'', and ``high-profile women''. Each of the three side panels represents
    one harmed subject. Panel 1 (top right) shows Naga Munchetty, annotated as an individual with gender value ``Woman'' marked incident-relevant because she was targeted with non-consensual deepfake nude images, and class value ``Celebrity'' also marked incident-relevant because her public figure status was used to commit fraud; age (49), language (English), and geography (UK) are marked not incident-relevant. Panel 2 (bottom right) shows Taylor Swift, annotated as an individual with gender value ``Women'' and class value ``High-profile'', both marked incident-relevant for the same reasons as Naga Munchetty. Panel 3 (bottom left) shows Megan Thee Stallion, annotated as an individual with gender value ``Women'' and class value ``High-profile'', both marked incident-relevant, with harms described as being targeted with deepfake nude images due to her gender and celebrity status.}
    \vspace{-12pt}
    \label{fig:report}
\end{figure}

For each incident report, the model applied the specified prompt. The model produced structured extractions aligned with the rubric, including harmed subjects, subject types, identity categories and values, and harm descriptions (see Figure \ref{fig:report} for an exemplary incident report and the corresponding extractions). The resulting dataset comprised 5,300 reports covering 1,200 unique incidents and 2,346 harmed subjects.

\subsubsection{Validating the Rubric Results.}
To assess extraction quality, we conducted a double-annotation exercise on a sample of 50 incident reports. Two annotators independently identified harmed subjects, assigned identity categories and values, and evaluated relevance using the rubric's counterfactual criteria. Agreement was high for harmed subject identification (92\% agreement, PABAK = 0.84) and causal relevance assessment (88\% agreement, PABAK = 0.76).\footnote{For these highly skewed distributions where both annotators predominantly selected ``Yes'', we report the Prevalence-Adjusted Bias-Adjusted Kappa (PABAK) rather than Cohen's $\kappa$, which underestimates agreement under class imbalance.} Identity category and value assignment showed substantial agreement (Cohen's $\kappa$ = 0.63), with annotators matching on 82\% of judgments. Most disagreements between the research team and the LLM arose in three situations.

First, when identity attributes were implied but not always explicitly stated in incident reports. For example, the LLM failed to  extract that Céline Dion is a woman \cite{AIID_Incident980}. Second, when the LLM mistook quoted claims about a subject for factual identity 
attributes. For example, it incorrectly inferred a political identity of  ``communist dictator'' for Kamala Harris from an AI-generated disinformation post such as quoted in the incident report \cite{AIID_Incident972}. Third, disagreements occurred when distinguishing between directly harmed subjects and affected bystanders. For example, in the Tay chatbot incident \cite{AIID_Incident6}, different members of the research team identified different harmed subjects, such as Black people or women, while the LLM labeled a broader group, such as U.S. social media users. After resolving disagreements through discussion between annotators, we computed LLM accuracy against the human gold standard. The LLM achieved 98\% accuracy on subject identification, 97\% accuracy on identity category value assignments (across 24 categories per subject), and 92\% accuracy on causal relevance judgments. To explore these disagreements, we examined the distribution of LLM misattributions across identity categories (i.e., cases where the LLM's identity category assignments differed from those of the annotators; Appendix~\ref{appendix:misattribution}, Figure \ref{fig:misattribution}). Misattributions were low across most categories (0--6\%), with the highest rate for gender (28\%). This rate reflects the first situation described above (i.e., the presence of implicit gender cues such as gender-coded titles or names that are inferred by human annotators but not by LLMs).

\subsection{Assessing the Relevance of Identity Categories in AI-Related Harms}
\label{subsec:step3}

Not all identity categories extracted from incident reports might be causally relevant to the harms described. Journalistic accounts often include personal details for context, and guidance on reporting on AI advises against reading such details as explanations for what went wrong, in order to avoid amplifying harm through misattribution \cite{UNESCO_AI_Journalism_2023}. We therefore assessed the relevance of each extracted identity category using the counterfactual evaluation defined in the rubric.

Specifically, for each identity category associated with a harmed subject, we analyzed the rubric's counterfactual component, which recorded two binary judgments: (CQ1) whether the incident occurred because the subject had this identity category, and (CQ2) whether the same incident would likely have occurred if the subject were identical in all respects except for this identity category. An identity category was retained only if it materially contributed to the AI system’s behavior (CQ1 = ``Yes'') and if changing the category would plausibly have altered the outcome (CQ2 = ``No''). Identity categories that did not meet this condition were removed from the subject record. If a harmed subject had no remaining relevant identity categories after this step, the subject was removed. If an incident had no remaining harmed subjects with relevant identity categories, the incident was removed.

To illustrate this step, consider an incident report describing the biased pregnancy prediction system deployed in Argentina, as introduced at the opening of this paper \cite{AIID_Incident188_2018}. The system generated individual risk scores from administrative data to guide follow-up interventions. Two identity categories are associated with the harmed subjects: gender (girls) and nationality (Argentine). For \emph{gender}, the LLM determined that the incident plausibly occurred because the subjects were girls (CQ1 = ``Yes''), and that the outcome would likely have differed had the subjects been boys (CQ2 = ``No''), since girls were singled out for risk scoring while boys were excluded entirely. The gender category was therefore retained. For \emph{nationality}, the LLM determined that the incident did not occur because the subjects were Argentine (CQ1 = ``No''), and that the same outcome would likely have occurred had the girls held a different nationality (CQ2 = ``Yes''). Because nationality did not plausibly influence the system’s behavior, this category was removed from the subject record.

After applying this relevance filtering, the resulting dataset comprised 711 unique incidents and 1,513 harmed subjects. This filtered dataset forms the basis for the analyses reported in the following section.

\subsection{Comparing Relevant Identity Categories of Harmed Subjects and Their Intersections Across AI Incidents}
\label{subsec:step4}

Using the filtered dataset, we address our two research questions by defining appropriate metrics and applying complementary quantitative and qualitative methods. We first analyze identity categories and values individually to assess their prevalence and associated harms (RQ\textsubscript{1}), and then examine their intersections (RQ\textsubscript{2}).
\medskip

\noindent{\emph{RQ\textsubscript{1}:} Which identity categories and their corresponding values appear most often in AI incidents reported in the news, and what kinds of harm are linked to them?} \\
\noindent We answered this question both quantitatively and qualitatively. Quantitatively, we defined 2 metrics: 

\begin{enumerate}[leftmargin=*]
    \item \emph{Category-level prevalence.} 
    To determine which identity categories are most represented in AI harm incidents, we calculated the prevalence of each category as:
    
    \begin{equation}
        \text{prevalence}_{\text{c}_i} = \frac{n_{\text{c}_i}}{N}
        \label{eq:category_prevalence}
    \end{equation}
    
    where $n_{\text{c}_i}$ is the number of incidents involving at least one subject of  identity category $c_i$ (e.g., gender, race), and $N$ is the total number of incidents in our dataset.
    
    \item \emph{Value-level prevalence.} 
    To examine which specific identity values within each category are most frequently associated with harm, we calculated:
    
    \begin{equation}
        \text{prevalence}_{\text{v}_i} = \frac{n_{\text{v}_i}}{N}
        \label{eq:value_prevalence}
    \end{equation}
    
    where $n_{\text{v}_i}$ is the number of incidents involving at least one subject with identity value $v_i$ (e.g., female for gender, Black for race).
\end{enumerate}

Qualitatively, we thematically analyzed harm examples associated with specific identities. We followed a four-step iterative coding process~\cite{saldana2015coding, miles1994qualitative, mcdonald2019reliability, braun2006thematic}. First, two coders from the research team independently coded a random sample of 50 incident reports by writing down the AI behavior, the identity categories and their specific values it targeted, and the resulting harm (e.g., behavior: AI generated sexualized images; identities: gender --- women, age --- girls; harm: sexual objectification). We discussed these codes among all authors, and refined them into a shared codebook. Second, the same two coders applied this codebook independently to the full dataset, and resolved any disagreements through discussion. Third, we reviewed the codes and clustered AI actions describing similar harms into a single pattern (e.g., sexualizing women and girls; mismatching people of color). Fourth, we refined these clusters into six final patterns, which we report with representative incidents in \S\ref{subsec:results_rq1}.

\medskip
\noindent{\emph{RQ\textsubscript{2}:} How do intersecting identity categories affect the likelihood and types of harm caused by AI systems?} \\
\noindent We answered this question both quantitatively and qualitatively. Quantitatively, we defined 2 metrics:

\begin{enumerate}[leftmargin=*]
    \item \emph{Intersectional score.} 
    To identify which combinations of identity categories co-occur most frequently in AI harm incidents, we calculated the joint prevalence of each category pair as:

    \begin{equation}
        \text{intersectional\_score}_{\text{c}_1, \text{c}_2}
        = \frac{n_{\text{c}_1, \text{c}_2}}{N}
    \end{equation}
    
    where $n_{\text{c}_1, \text{c}_2}$ is the number of incidents involving both identity categories (e.g., gender and age) and $N$ is the total number of incidents. We visualize co-occurrences in a matrix to reveal intersectional harm patterns.

\item \emph{Amplification score.}
    To assess whether for particular intersecting categories, specific identity value combinations are disproportionately represented among AI-harmed subjects beyond what would be expected by chance, we computed a conditional amplification score:

    \begin{equation}
    \label{eq:amplification_score}
    \text{amplification\_score}_{\text{v}_1, \text{v}_2}
    = \frac{n_{\text{v}_1, \text{v}_2}}{\mathbb{E}[n_{\text{v}_1, \text{v}_2}]}
    \end{equation}
    
    Here,  $n_{\text{v}_1, \text{v}_2}$ is the number of incidents involving both identity category values $v_1$ and $v_2$ (e.g., a Black woman), and $\mathbb{E}[n_{\text{v}_1, \text{v}_2}]$ is the expected number of incidents under the assumption that $v_1$ and $v_2$ are independent:

    \begin{equation}
       \mathbb{E}[n_{\text{v}_1, \text{v}_2}] = \frac{n_{\text{v}_1} \times n_{\text{v}_2}}{N}
    \end{equation}
    
    A score greater than 1 indicates that the value combination occurs more frequently than expected by chance, reflecting amplified vulnerability to AI harm; a score below 1 indicates lower-than-expected occurrence.

\end{enumerate}

Qualitatively, we thematically analyzed harms associated with  intersecting identity values. First, we selected harms involving multiple identity categories and values (e.g., harm: sexual objectification; identities: gender --- women \emph{and} age --- girls). Second, we coded and grouped these harms into patterns at each intersection. Third, we clustered patterns across intersections into five intersectional harm themes, which we report with 
representative incidents in \S\ref{subsec:results_rq2}.
\section{Results}
\label{sec:results}

We report results in two steps, first examining identity categories and values in isolation (\S\ref{subsec:results_rq1}), and then analyzing how intersecting identities shape the likelihood and forms of AI-related harm (\S\ref{subsec:results_rq2}).

\subsection{\emph{RQ\textsubscript{1}:} Which identity categories and their corresponding values appear most often AI incidents reported in the news, and what kinds of harm are linked to them?}
\label{subsec:results_rq1}

\begin{figure}[h!]
    \centering
    \includegraphics[width=0.89\linewidth]{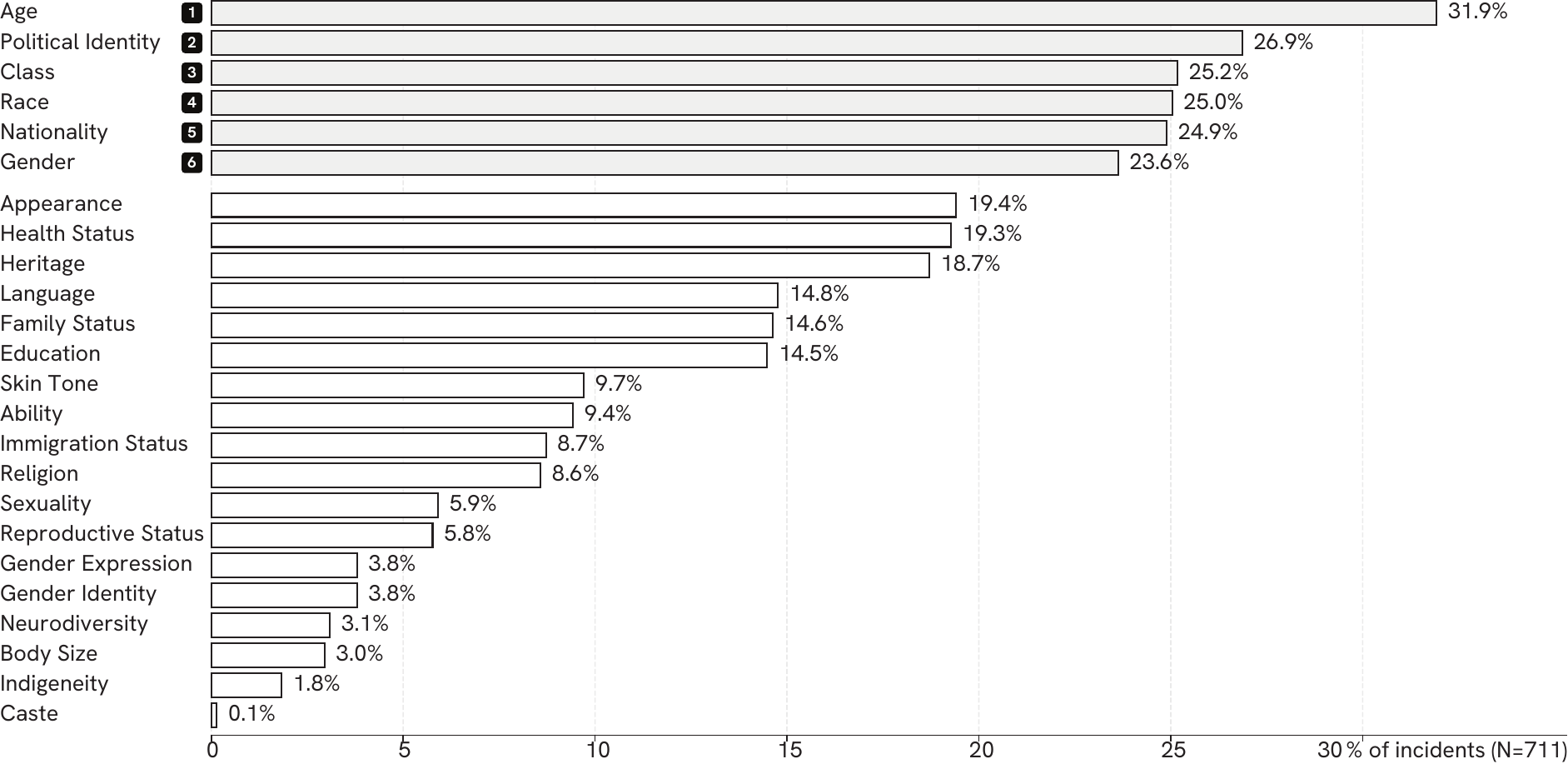}
    \caption{\textbf{Percentage of incidents in which the identity category was causally relevant to the incident (prevalence of the identity category as per Equation~\ref{eq:category_prevalence}).} The figure shows the percentage of incidents ($N = 711$) in which each identity category was identified as causally relevant to harm, counted once per incident; the six most prevalent categories (1--6) each appear in over 20\% of incidents. Age and political identity appear as frequently as race and gender, followed by class and nationality. Least frequently documented categories include body size, indigeneity, and caste.}
    \Description{A horizontal bar chart showing the prevalence of identity categories across AI incidents where identity was causally relevant to harm. The x-axis represents the percentage of incidents, ranging from 0 to 30 percent, and the y-axis lists identity categories. Percentages are calculated over 711 incidents, with each category counted at most once per incident. The six most prevalent identity categories, each appearing in more than 20 percent of incidents, are shown at the top and numbered one through six. These are: age (31.9 percent of incidents), political identity (26.9 percent), class (25.2 percent), race (25.0 percent), nationality (24.9 percent), and gender (23.6 percent). A second tier of moderately prevalent categories appears in approximately 15 to 20 percent of incidents, including skin tone (19.4 percent), ability (19.3 percent), immigration status (18.7 percent), religion (14.8 percent), sexuality (14.6 percent), and reproductive status (14.5 percent). Less frequently documented categories, each appearing in under 10 percent of incidents, include gender expression (9.7 percent), gender identity (9.4 percent), neurodiversity (8.7 percent), body size (8.6 percent), indigeneity (5.9 percent), caste (5.8 percent), appearance (3.8 percent), health status (3.8 percent), heritage (3.1 percent), language (3.0 percent), and family status (1.8 percent). Education is the least prevalent category, appearing in approximately 0.1 percent of incidents. Overall, the chart shows that age and political identity are as common as race and gender in documented identity-related AI harms, while categories such as body size, indigeneity, caste, and education are rarely reported.}
    \label{fig:categories}
\end{figure}

\textbf{Identity-related AI harms span all identity categories, with age and political identity appearing at rates comparable to race and gender.} 
We found 1,513 harmed subjects across 711 unique incidents. Among these, age emerges as the most prevalent identity category, appearing in 32\% of incidents, followed by political identity (27\%), class (25\%), race (25\%), nationality (25\%), and gender (24\%) (Figure \ref{fig:categories}). 

\begin{figure}[h!]
    \centering
    \includegraphics[width=0.89\linewidth]{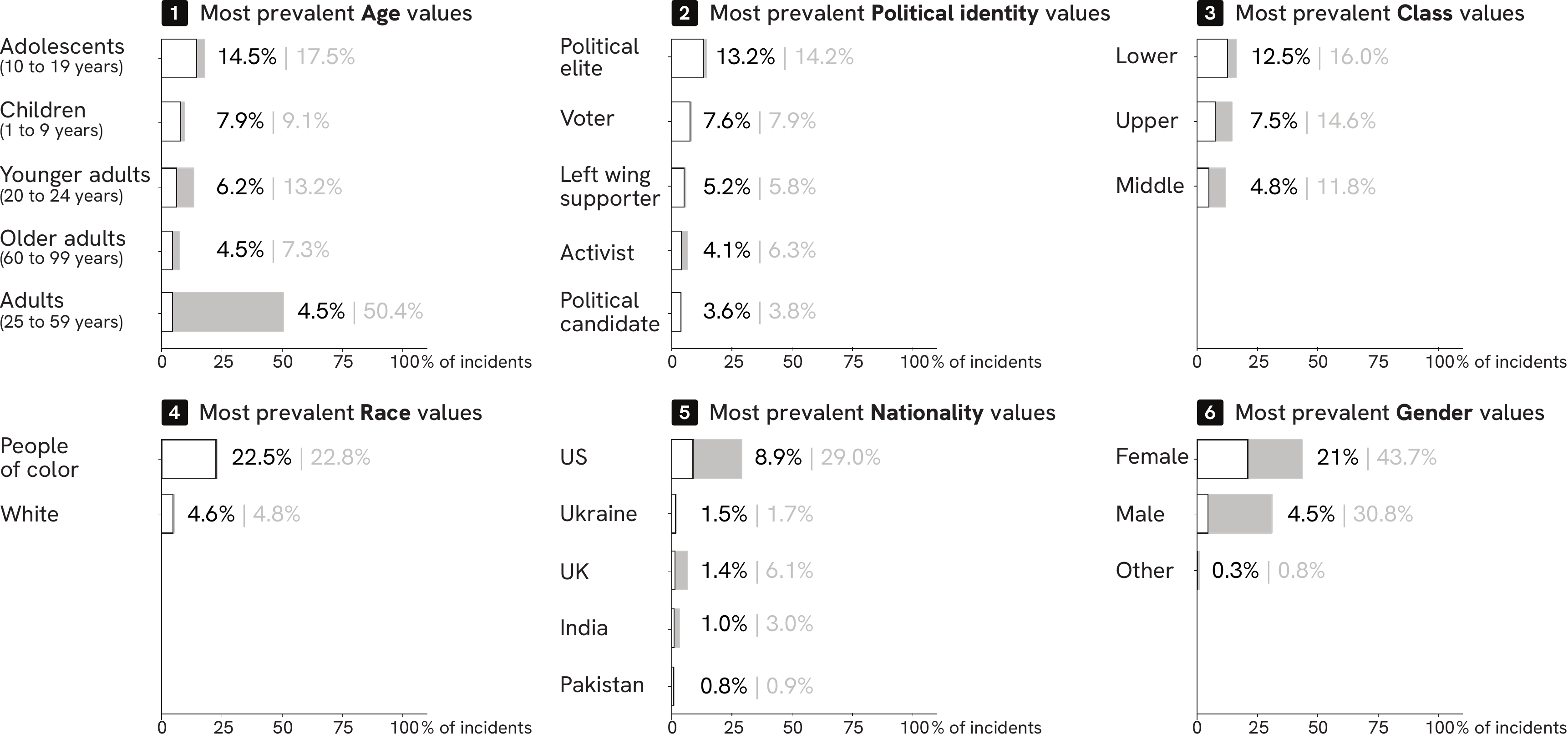}

    \caption{\textbf{Percentage of incidents for the most prevalent values within each of the six most prevalent categories (Figure \ref{fig:categories}).}
    Each panel shows the percentage of incidents in which a given identity value appears (prevalence of the identity value as per Equation~\ref{eq:value_prevalence}. Grey rectangles show overall prevalence across incidents, and white rectangles show prevalence when the identity value was causally relevant to the harm counted once per incident). Adolescents, political elites, individuals with lower class, people of color, U.S. nationals, and females appear most frequently within their respective categories.}
    
    \Description{A six-panel horizontal bar chart showing the most prevalent identity values within the six most prevalent identity categories in AI incidents with identity-related harm. Each panel corresponds to one identity category: age, political identity, class, race, nationality, and gender. For each identity value, two bars are shown: a grey bar indicating overall prevalence across incidents, and a white bar indicating prevalence when the identity value was causally relevant to harm, with each value counted at most once per incident. Percentages are shown on a 0 to 100 percent scale. 
    
    Panel 1 (Age): Adolescents aged 10–19 appear in 14.5 percent of incidents when causally relevant and 17.5 percent overall, making them the most prevalent age group. Children aged 1–9 appear in 7.9 percent causally and 9.1 percent overall. Younger adults aged 20–24 appear in 6.2 percent causally and 13.2 percent overall. Older adults aged 60–99 appear in 4.5 percent causally and 7.3 percent overall. Adults aged 25–59 appear in 4.5 percent causally but dominate overall prevalence at 50.4 percent.
    
    Panel 2 (Political identity): Political elites are the most prevalent, appearing in 13.2 percent of incidents when causally relevant and 14.2 percent overall. Voters appear in 7.6 percent causally and 7.9 percent overall. Left-wing supporters appear in 5.2 percent causally and 5.8 percent overall. Activists appear in 4.1 percent causally and 6.3 percent overall. Political candidates appear least frequently, at 3.6 percent causally and 3.8 percent overall.

    Panel 3 (Class): Individuals identified as lower class are most prevalent, appearing in 12.5 percent of incidents when causally relevant and 16.0 percent overall. Upper-class individuals appear in 7.5 percent causally and 14.6 percent overall. Middle-class individuals appear least frequently, at 4.8 percent causally and 11.8 percent overall.

    Panel 4 (Race): People of color appear in 22.5 percent of incidents when race is causally relevant and 22.8 percent overall, making them the dominant race category. White individuals appear in 4.6 percent causally and 4.8 percent overall.

    Panel 5 (Nationality): U.S. nationals are the most prevalent nationality, appearing in 8.9 percent of incidents causally and 29.0 percent overall. Ukrainian nationals appear in 1.5 percent causally and 1.7 percent overall. UK nationals appear in 1.4 percent causally and 6.1 percent overall. Indian nationals appear in 1.0 percent causally and 3.0 percent overall. Pakistani nationals appear least frequently, at 0.8 percent causally and 0.9 percent overall.

    Panel 6 (Gender): Females are the most prevalent gender category, appearing in 21.0 percent of incidents when causally relevant and 43.7 percent overall. Males appear in 4.5 percent causally and 30.8 percent overall. Other genders appear rarely, at 0.3 percent causally and 0.8 percent overall.

    Overall, the figure shows that within each top identity category, a small number of identity values account for most causally relevant harms, and that values with high overall prevalence do not always correspond to high causal relevance.}
    \label{fig:values}
\end{figure}

\smallskip
\noindent
\textbf{Within identity categories, AI harms concentrate on those who are structurally exposed rather than numerically dominant.} In age-based incidents, adolescents (14\%) and children (8\%) appear more frequently than adults (5\%), despite comprising a smaller share of the populations affected by many AI systems (Figure~\ref{fig:values}). Political identity illustrates this pattern clearly: harms concentrate on structurally exposed roles such as political elites (13\%), voters (7\%), activists (4\%), and party candidates (4\%), rather than on larger ideological groups like left-wing individuals (5\%). Class-based harms disproportionately affect lower-class subjects (12\%). Race-related harms primarily involve people of color (22\%), while incidents involving white subjects are far less common (5\%). Gender-related harms overwhelmingly affect females (21\%), compared to males (4\%).

\smallskip
\noindent
\textbf{Single identity category AI harms recur through a small set of mechanisms by which systems act on one socially salient value and scale its consequences.} The thematic analysis shows that these harms follow six recurring patterns in how AI systems act on a single identity value, namely by \emph{sexualizing}, \emph{steering}, \emph{matching}, \emph{inferring}, \emph{gating}, or \emph{manipulating} individuals. \emph{Sexualizing} is most visible in harms affecting women and girls, where generative systems are used to produce and circulate non-consensual sexual imagery such as deepfakes. This turns femininity into a persistent site of reputational and psychological attack (in incidents \cite{AIID_Incident314, AIID_Incident610, AIID_Incident874}), with especially acute effects when such material circulates in school or peer contexts (see incident \cite{AIID_Incident717}). 

\noindent\emph{Steering} operates most clearly in harms affecting men and boys, where recommender systems systematically funnel them toward misogynistic or extremist content, shaping norms around dominance, hostility, and political violence rather than simply reflecting pre-existing preferences (see incidents \cite{AIID_Incident263, AIID_Incident300}). \emph{Matching} becomes harmful where biometric identification systems misrecognize racialized individuals: facial recognition technologies disproportionately misidentify Black people and convert technical error into wrongful stops or arrests (see incidents \cite{AIID_Incident74, AIID_Incident244, AIID_Incident288, AIID_Incident592}). \emph{Inferring} produces distinct harms when AI systems attempt to predict or reveal sexual orientation, exposing individuals to risks of outing, targeting, and physical danger without consent (see incidents \cite{AIID_Incident167, AIID_Incident431}). \emph{Gating} emerges where access to work or services is conditioned on rigid identity verification, as in biometric systems that assume binary gender presentation and treat change as anomaly, leading to automated exclusion and income loss for gender-diverse people (see incident \cite{AIID_Incident396}). Finally, \emph{manipulating} captures harms tied to political identity, where generative media and personalization systems fabricate imagery or narratives attributed to specific political groups, shaping beliefs by exploiting identity-based trust (see incidents \cite{AIID_Incident650, AIID_Incident862}).

\subsection{\emph{RQ\textsubscript{2}:} How do intersecting identity categories affect the likelihood and types of harm caused by AI systems?}
\label{subsec:results_rq2}

\begin{figure}[t]
    \centering
    \includegraphics[width=0.635\linewidth]{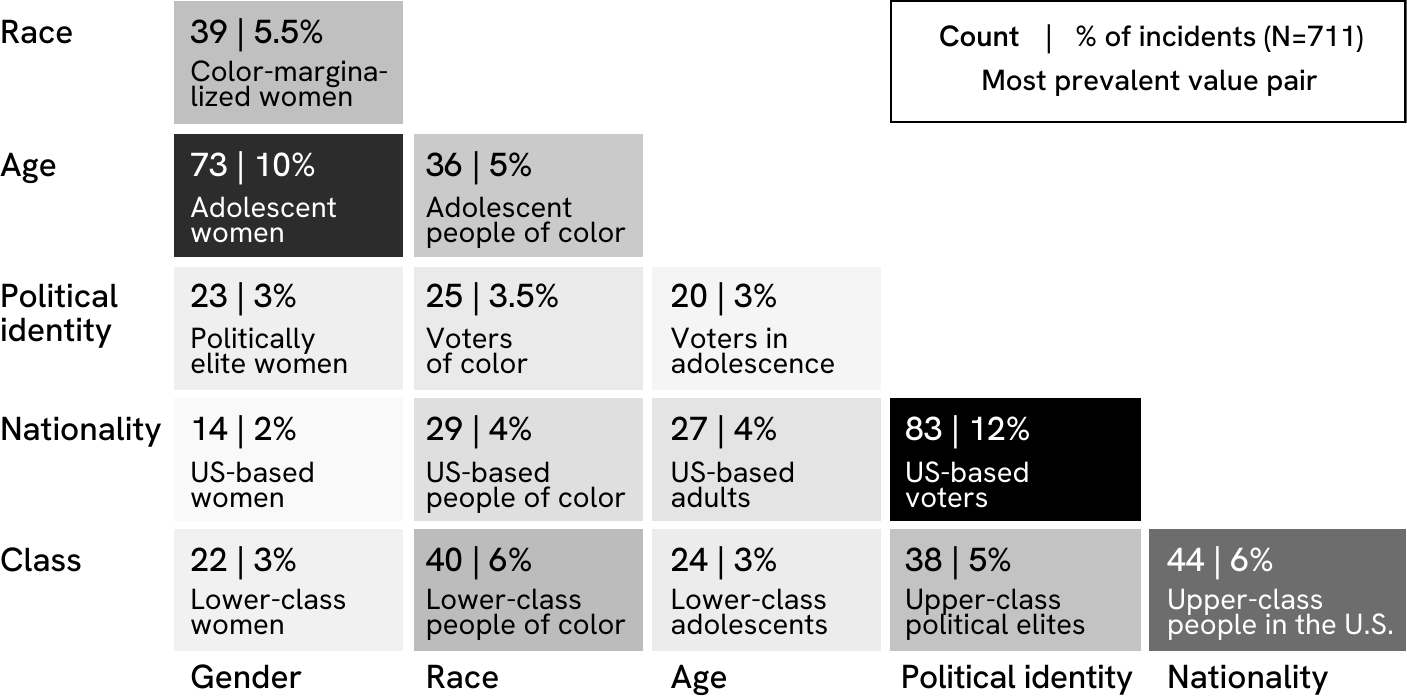}
    \caption{\textbf{Prevalence of intersecting identity categories in AI incidents with identity-related harm.} Each cell shows the number and share of incidents ($N = 711$) involving subjects with both categories, counted once per incident. Darker shading indicates higher prevalence. The most frequent intersections involve nationality and political identity, age and gender, and nationality and class.} 
    
    \Description{A heatmap showing the prevalence of intersecting identity categories in AI incidents with identity-related harm. Both the rows and columns list six identity categories: race, age, political identity, nationality, class, and gender. Each cell represents incidents in which harmed subjects possessed both identity categories, counted once per incident, with a total of 711 incidents. Each cell displays both the number of incidents and the corresponding percentage of all incidents. Darker shading indicates a higher prevalence of that intersection.

    The most prevalent intersection appears between nationality and political identity, with U.S.-based voters involved in 83 incidents, representing 12 percent of all incidents. Other highly prevalent intersections include age and race, such as adolescents of color (36 incidents, 5 percent), and class and race, such as lower-class people of color (40 incidents, 6 percent).
    
    Several intersections involving gender are also common. These include women of color (39 incidents, 5.5 percent), politically elite women (23 incidents, 3 percent), lower-class women (22 incidents, 3 percent), and U.S.-based women (14 incidents, 2 percent).
    
    Age intersects frequently with multiple categories, including voters in adolescence (20 incidents, 3 percent), lower-class adolescents (24 incidents, 3 percent), and adolescents of color. Nationality similarly intersects with race, class, and age, including U.S.-based people of color (29 incidents, 4 percent), U.S.-based adults (27 incidents, 4 percent), and upper-class people in the United States (44 incidents, 6 percent).
    
    Overall, the heatmap shows that AI-related identity harms most often involve overlapping categories tied to nationality, political identity, age, race, and class, rather than single identity categories in isolation.}
    \vspace{-10pt}
    \label{fig:heatmap}
\end{figure}

\smallskip
\noindent
\textbf{Intersectional harms are not evenly distributed across category combinations.} Instead, a few pairings recur frequently across AI incidents (Figure \ref{fig:heatmap}). The most common intersection is nationality and political identity (12\%), followed by age and gender (10\%) and nationality and class (6\%).

\begin{figure}[t]
    \centering
    \includegraphics[width=\linewidth]{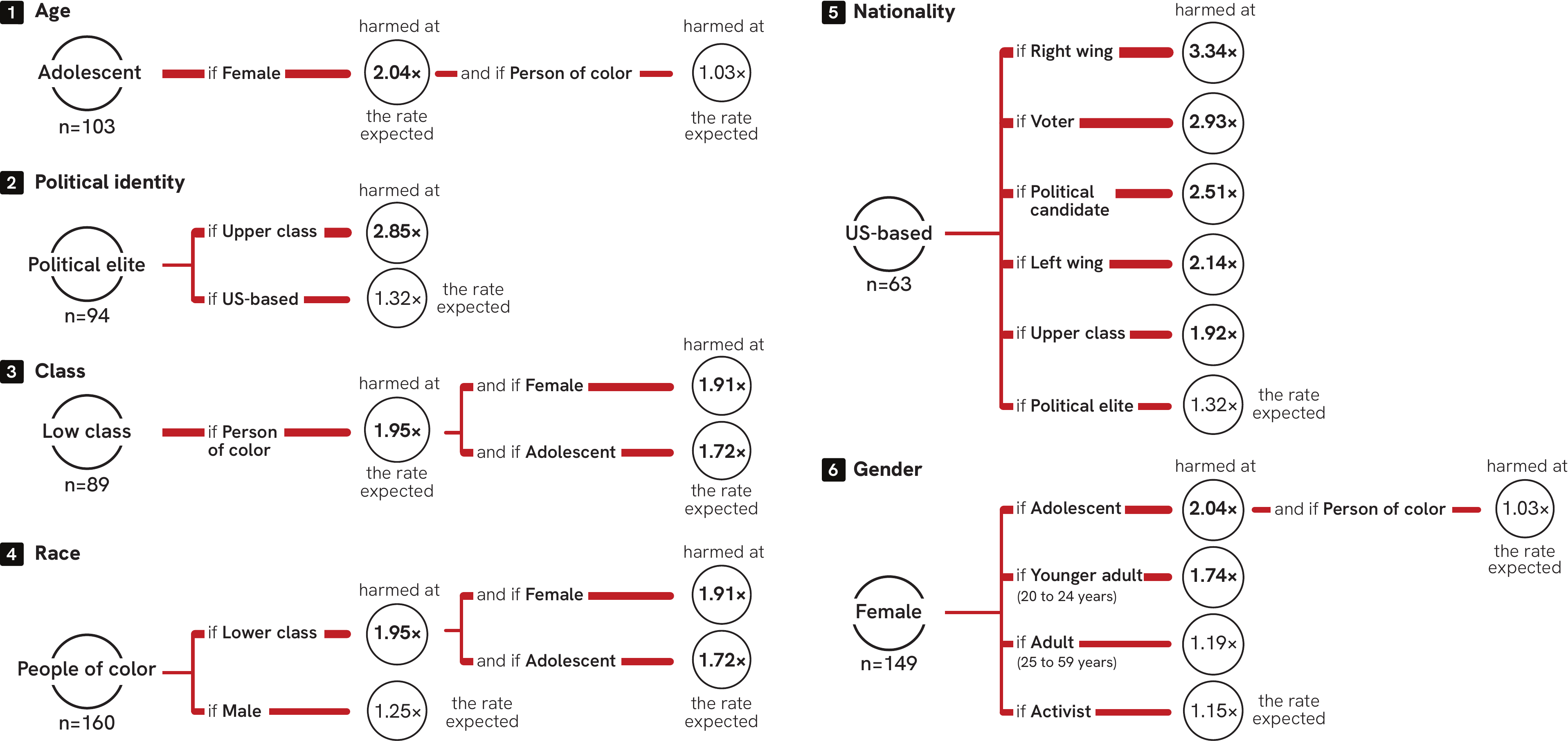}
    \caption{\textbf{The amplification scores for the six most prevalent categories in Figure \ref{fig:categories}.} Line width and labels inside the circles with an $\times$ amplification factor report the amplification score between a pair of two identity values, calculated as per Formula \ref{eq:amplification_score}. Scores greater than 1 indicate that a given intersection occurs at \emph{$\times$}-times the rate expected under independence of the two categories. The strongest amplification is observed for \texttt{US} and \texttt{right-wing} voters (3.34$\times$--2.93$\times$): from point~5, the observed number of harmed incidents involving subjects who are \texttt{US} and \texttt{right-wing},
    \(n_{\texttt{US},\,\texttt{right-wing}}\), is \(3.34\times\) the expected number under the assumption that
    \texttt{US} and \texttt{right-wing} are independent. This value pair is followed by intersections of political elites with upper-class status (2.85$\times$), and adolescent age with female gender (2.04$\times$).}
    \Description{A node–link diagram showing amplification effects between intersecting identity values drawn from the six most prevalent identity categories in AI-related incidents reported in the news. Each identity value is represented as a labeled circle with the number of harmed subjects shown next to it. Lines connect pairs of identity values, and the thickness of each line corresponds to an amplification score. A numeric label inside a circle along each line shows the amplification factor, indicating how many times more often the intersection occurs than expected if the two identity categories were independent. Values greater than 1 indicate amplified harm.

    Nationality: U.S.-based subjects show the strongest amplification effects when intersecting with political identities. U.S. and right-wing voters have the highest amplification, occurring between 3.34 and 2.93 times the expected rate. Other amplified intersections include U.S. with political candidates (2.51 times), left-wing supporters (2.14 times), upper-class status (1.92 times), and political elites (1.32 times).
    
    Political identity: Political elites show strong amplification when intersecting with upper-class status (2.85 times) and moderate amplification with U.S.-based nationality (1.32 times).
    
    Age: Adolescents show amplified harm when intersecting with female gender (2.04 times) and near-expected rates when intersecting with being a person of color (1.03 times).
    
    Gender: Female subjects show amplification when intersecting with adolescent age (2.04 times), younger adults aged 20–24 (1.74 times), adult age 25–59 (1.19 times), and activist political identity (1.15 times).
    
    Class: Lower-class status shows amplified harm when intersecting with being a person of color (1.95 times), female gender (1.91 times), and adolescent age (1.72 times).
    
    Race: People of color show amplified harm when intersecting with lower-class status (1.95 times), female gender (1.91 times), and adolescent age (1.72 times), with a smaller amplification when intersecting with male gender (1.25 times).
    
    Overall, the diagram shows that intersections involving nationality and political identity exhibit the strongest amplification of harm, while intersections involving age, gender, race, and class also frequently amplify harm beyond what would be expected from single identity categories alone.}
    \vspace{-11pt}
    \label{fig:nodemap}
\end{figure}

\smallskip
\noindent
\textbf{Female gender, adolescence, lower class status, and political elite roles stand out for sharply amplifying harm when they intersect, often appearing at least twice as often as expected.} Intersectional effects are not merely additive. Several identity value combinations appear in incidents far more often than would be expected based on their individual prevalence alone (Figure~\ref{fig:nodemap}). The intersection of female gender with adolescent age appears at more than twice the rate expected under independence of the two categories. A contrasting gender and age pattern emerges for male subjects. Male gender intersects not with adolescent age but with adulthood, appearing nearly three times more often than expected. This shows that amplification also arises when socially dominant categories align. Socioeconomic and racial intersections also show strong amplification. Lower class status intersecting with being a person of color appears at nearly twice the expected rate, illustrating how marginalized identities compound one another. Upper class status intersecting with political elite membership appears nearly three times more often than expected, similarly amplifying already dominant positions. By contrast, several identity values that are common in isolation show little or no amplification when intersected. For example, male gender intersecting with race yields amplification score close to one, indicating no elevated harm. These patterns show that intersectional amplification emerges at both ends of the power spectrum, compounding harm for marginalized groups like women and adolescents, and concentrating risk even among dominant, highly visible groups such as political elites.

\smallskip
\noindent
\textbf{Intersectional AI harms arise through processes that link AI systems to social power.} We identified five dominant intersectional harm themes (Table~\ref{tab:intersections}, Appendix \ref{appendix:incident_table}).

First, intersectional harms cluster around \emph{algorithmic suspicion and criminalization}. These harms are driven by risk-scoring, fraud-detection, and classification systems in welfare administration, immigration control, and criminal justice. They disproportionately affect low-income, racialized, and migrant families. For example, in the Dutch childcare benefits system, dual-nationality families were flagged as fraudulent, leading to debt and family separation (see IDs 40, 101, and 335 in Table~\ref{tab:intersections}). In this theme, AI systems operationalize historically racialized and classed logics of suspicion, reflecting racism, classism, and nativism.

Second, a recurring pattern of \emph{sexualized exploitation} arises from generative media systems and platform-scale recommender infrastructures. They disproportionately affect adolescent girls and young women, particularly those from lower-class, Indigenous, or migrant communities. Harms include non-consensual deepfakes, ``nudification'', grooming, and invasive prediction of reproductive futures. An opening example of this paper concerns a predictive analytics system in Argentina that classified low-income Indigenous girls as at risk'' of adolescent pregnancy \cite{AIID_Incident188_2018}. These girls were often subjected to intensive data collection and surveillance linked to access to social services (see IDs 188, 904, and 924 in Table~\ref{tab:intersections}). Here, AI systems amplify sexism and ageism, reinforcing patriarchal control over women's bodies.

Third, intersectional harms emerge through \emph{environmental violence}. These harms are produced by AI-enabled technical infrastructure and allocation systems in energy, urban planning, and healthcare. They disproportionately burden racialized and lower class communities with children. A clear example is the siting of xAI's supercomputer facility in South Memphis, which concentrated pollution in historically Black neighborhoods, increasing health risks for children and families without local consent or governance input (see ID 1144 in Table~\ref{tab:intersections}). In this theme, AI harms reflect environmental racism and classism.

Fourth, we identify \emph{political manipulation and democratic erosion} enabled by AI-generated media, synthetic personas, and personalization systems. These systems operate in electoral and political communication contexts. They target young voters and racialized communities with tailored disinformation. AI-generated images falsely depicting Black political support illustrate this dynamic (see IDs 202, 650, and 972 in Table~\ref{tab:intersections}). These cases exploit forms of social power such as ageism, racism and nationalism to shape participation and public perception.

A final theme involves \emph{militarized and state violence} enabled by AI. These harms are driven by surveillance, targeting, and threat-classification systems in policing, security, and warfare. They disproportionately affect civilians at the intersections of nationality, race, and political identity. For example, the Lavender AI system used during Israel's war in Gaza (see ID 672 in Table~\ref{tab:intersections}) classified Palestinian men as suspected militants and identified their family homes as targets for airstrikes, often with minimal human review, resulting in large-scale civilian deaths. These cases reflect extreme forms of nationalism, racism, and militarism where automated classification directly mediates life-and-death outcomes.
\section{Discussion}
\label{sec:discussion}

This study examines identity-related AI harms at a scale and breadth rarely achieved in prior work. Combining large-scale AI incident analysis with an intersectional rubric and thematic analysis, we surface both established and underexplored harm patterns. We interpret these findings in relation to prior literature (\S\ref{subsec:inline_results}), discuss contributions to intersectionality-informed AI research (\S\ref{subsec:challenging_results}), and outline implications, limitations, and future directions (\S\ref{subsec:implications}, \S\ref{subsec:limitations}).

\subsection{In-line with Previous Literature}
\label{subsec:inline_results}

Our findings align with prior research in two key areas: (1) persistent disparities along race and gender lines in AI harm, and (2) the disproportionate targeting of structurally vulnerable groups.

First, we replicate well-documented racial and gender disparities in AI harms. Prior work showed that algorithmic systems disproportionately misrepresent or disadvantage racialized and feminine-coded individuals \cite{HillCollins2002, GenderShades_2018}. Our findings reinforce this pattern, with AI-related harm incidents disproportionately involving people of color and women.

Second, we confirm that AI harms concentrate on specific social groups rather than on numerically dominant populations. Low-income families experience disproportionate harm from automated decision-making systems that govern access to welfare, where risk scores can worsen financial instability \cite{eubanks2018automating}. Racialized and minority groups are disproportionately affected by surveillance and classification systems that intensify unequal monitoring \cite{benjamin2019raceaftertech}. Children are frequently harmed by AI systems in digital environments despite having limited capacity to opt out from automated decisions. These patterns support critiques that fairness frameworks often overlook how harm is shaped by power, governance, and unequal exposure to institutional systems \citep{hoffmann2019fairness}. Our results reinforce this critique by showing elevated harm among children and low-income subjects, indicating that AI systems amplify existing social vulnerabilities.

\subsection{Contributions to Previous Literature}
\label{subsec:challenging_results}

Our findings challenge prevailing assumptions in responsible AI research in two key ways: (1) by broadening the landscape of identity categories implicated in AI risk assessments beyond the field's dominant focus on race and gender, and (2) by showing that intersectional harms arise not only from the compounding of marginalizations but also through concentrated exposure at the high-end spectrum of power.

First, we find that AI harms related to age and political identity occur as frequently as those based on race and gender. This challenges the dominant framing in algorithmic fairness research, which has historically focused on race and gender as the primary axes of harm \citep{GenderShades_2018,noble2018algorithms}. Age has been largely neglected as a core dimension of algorithmic discrimination. Researchers have recently called attention to AI ageism \cite{Stypinska2022}, where older adults face exclusion, misclassification, or stereotyping in AI systems across domains like employment and health care \cite{AffectRecognition_2022}. Age-based vulnerability also extends to younger populations, as AI systems embedded in children's toys, tutors, and media ecosystems deliberately personalise learning, play, and social interaction, raising age-specific risks of reduced exposure to diverse experiences, dependency on compliant AI companions, and early immersion in highly personalised algorithmic environments \cite{EconomistAIRewiringChildhood_2025}. Political identity has emerged as a significant axis of vulnerability. \citet{Peters2022} argue that political bias in algorithmic systems can operate like racial or gender bias but with fewer social or institutional safeguards. Our findings support this concern, showing that activists, candidates, and voters are regularly harmed by AI-mediated impersonation.

Second, we find that intersectional harms are not confined to the most marginalized groups. Previous work on intersectionality in AI has appropriately focused on how intersecting systems of oppression compound risk for subjects at the margins such as women of color or low-income migrants \cite{AlgorithmicDisadvantage_2024}. Our findings do not contradict this focus but extend it by showing that intersectional harms can also arise through concentrated exposure at positions of institutional visibility and power. Political elites are one such example. In our dataset, they frequently appear in incidents where AI systems are used to manipulate public narratives that undermine trust and participation \cite{restofworld2024ai}. To see how this occurs, consider how Cara Hunter was targeted by a deepfake pornographic video during an election campaign, an incident widely reported to have nearly ended her political career \cite{hunter2025deepfake}. Similarly, older affluent individuals can face exclusion through age-based algorithmic profiling in domains such as credit scoring and hiring, where automated systems treat age itself as a risk or liability, overriding otherwise advantaged class positions and professional credentials \cite{Stypinska2022}.

\subsection{Implications}
\label{subsec:implications}
Our findings show that addressing AI harms requires moving beyond risk occurring for specific identities in isolation toward identifying and mitigating risks occurring at specific identity intersections. In practice, this means that intersectionality can be used to: (1) identify which groups are structurally exposed during system design; (2) evaluate systems against intersection-specific harm patterns at deployment; and (3) monitor how these harms emerge and accumulate over time. We elaborate these implications across three phases of the AI lifecycle: design, deployment, and monitoring.

\smallskip
\noindent
\textbf{Design Phase}. The rubric introduced in this work (Appendix \ref{appendix:categories_list}, \ref{appendix:rubric_prompt}) can support early risk anticipation by moving beyond single-axis notions of intended users. Design teams can use it internally to stress-test assumptions about ``typical'' users and to identify which identity intersections are likely to be structurally exposed within a given domain. For example, in public-sector systems, categories such as language or migration status may be more consequential than the race–gender combinations emphasized in many Western fairness frameworks. At the same time, our findings caution against uncritical use of identity-based personas during design. When applied superficially, intersectionality can reinforce stereotypes rather than reveal how institutional systems unevenly affect different groups.

\smallskip
\noindent
\textbf{Deployment Phase}. 
Our analysis informs guidance for audits, procurement, and regulatory oversight. For example, the concentration of non-consensual image generation involving adolescent girls calls for heightened scrutiny of AI tools used in schools and messaging apps. Similarly, harms involving adult men being algorithmically steered toward misogynistic or extremist content indicate that recommender systems should be assessed for how they shape behavior across specific gender–age cohorts, rather than solely through aggregate engagement metrics. Procurement teams and regulators can use these empirically observed harm patterns to request documentation on how vendors identify and mitigate intersection-specific risks. Such evaluations should consider not only incident frequency but also harm severity and downstream impact. As we show in our work, some intersections appear less often yet carry disproportionate consequences such as long-term reputational damage.

\medskip
\noindent
\textbf{Monitoring Phase}. 
Our findings underscore the value of participatory end-user auditing for detecting intersectional harms that emerge or intensify post-deployment. Many harms become visible only through sustained use and lived experience, particularly for structurally exposed groups. The rubric can support participatory audits by providing a shared structure for documenting identity-relevant harms and their concrete impacts, without requiring technical expertise. Integrating such audits into ongoing monitoring helps surface rare but severe, cumulative, or normalized harms that may remain invisible to developer-centric metrics such as fairness benchmark performance.

\subsection{Limitations and Future Work}
\label{subsec:limitations}

Our work comes with four main limitations that point to directions for future research.

First, it relies on publicly documented AI incidents shaped by news media visibility and disclosure practices. This introduces reporting bias: harms affecting less visible populations or occurring in closed systems are likely underrepresented~\cite{DecodingIncidents_2024}. Future work could incorporate regulatory filings or whistleblower reports.

Second, incident reports primarily describe individuals or groups directly affected by AI uses and rarely document ripple effects on other populations. Our analysis therefore focuses on explicitly named subjects and cannot capture indirect harms across social, organizational, or institutional contexts. Future work could extend incident reporting frameworks to capture secondary harms and affected stakeholders \cite{taxonomyCDI2024}, or use strategic foresight to anticipate them \cite{systemicRisks2026}.

Third, the incident reports are skewed toward U.S.-based subjects, limiting generalizability to regions where AI harms may be underreported, differently framed, or embedded in distinct sociopolitical contexts. As a result, the prevalence and amplification patterns reflect locally documented harms rather than the global distribution of AI-related harm. Future work should prioritize incident collection in non-U.S. and non-English-speaking contexts and examine how intersectional harm manifests across different regulatory, cultural, and infrastructural settings.

Fourth, our use of LLMs enables scalable analysis but reflects the interpretive frame of the rubric and model responses. Although we applied counterfactual filtering and validation to improve reliability, future work could incorporate complementary approaches such as expert or community-based annotation, to surface alternative interpretations.
\section{Conclusion}
\label{sec:conclusion}

This paper delivers an empirical analysis to date of intersectional AI harms by examining 5,300 reports across 1,200 incidents. It identifies recurring intersectional harm patterns that show how AI systems amplify both structural vulnerability and institutional visibility across diverse domains. These contributions broaden the empirical foundations of responsible AI research and demonstrate why AI risk assessment must move beyond narrow fairness categories to address the identity configurations most commonly harmed in practice.
\clearpage
\section{Endmatter Statements}
\label{sec:statements}

\subsection{Generative AI Usage Statement}
We used generative AI tools for three tasks: data analysis, manuscript development, and accessibility support, as described below.

\smallskip
\noindent
\textbf{Use in Data Analysis.} We used a state-of-the-art, commercially available LLM, OpenAI GPT-5.1, via the API, to systematically extract information about harmed subjects, identity categories, and associated harms from AI incident reports in the AI Incident Database. The model was used to apply a structured rubric through prompt-based batch processing, including counterfactual relevance assessment, as described in the Methods sections (\S\ref{subsec:step2} and \S\ref{subsec:step3}). All model outputs were reviewed by the authors and filtered according to predefined criteria before inclusion in the final dataset.

\smallskip
\noindent
\textbf{Use in Manuscript Development.} We used another LLM, Google Gemini 3 Pro, via the chat interface to support grammar checking and stylistic edits in selected paragraphs of the manuscript. These tools were not used to generate original research content, results, or interpretations.

\smallskip
\noindent
\textbf{Use in Accessibility Support.} We also used Google Gemini 3 Pro via the chat interface to assist with drafting initial alt-text descriptions for figures and tables. These descriptions were subsequently reviewed and revised by the authors.

\subsection{Ethical Considerations Statement}
Our work raises four ethical considerations.

First, our reliance on news-based incident reports means that our findings reflect which harms are deemed newsworthy rather than which occur most frequently. Harms affecting communities with less media visibility may therefore be underrepresented. We caution against interpreting the absence of certain identity configurations as evidence that such harms do not occur.

Second, categorizing individuals into discrete identity values (e.g., ``Black'', ``woman'', ``lower-class'') risks reifying categories that are fluid and socially constructed. We adopted this approach to enable systematic analysis while recognizing that it can obscure within-group heterogeneity.

Third, the use of LLMs to assist in coding introduces potential biases, as models may reflect stereotypes embedded in their training data. We mitigated this risk through a structured rubric and manual validation, but acknowledge that automated identification of sensitive attributes is inherently imperfect.

Fourth, our analysis emphasizes the frequency and co-occurrence of identity categories in documented incidents, but frequency alone does not capture the seriousness or impact of harm. Some harms may appear less often in incident data yet carry disproportionate social, psychological, or material consequences. Our results should therefore not be interpreted as ranking harms by importance or severity.

\clearpage

\bibliographystyle{ACM-Reference-Format}
\bibliography{main}

\appendix
\clearpage
\section*{Appendix}

\section{Criteria for Selecting the Source of AI Incident Reports}
\label{appendix:criteria}

We defined three criteria for selecting a source of AI incident reports: (C1) whether the source preserves original materials such as serious incident reports \cite{EUAIAct2024}, (C2) whether those materials contain sufficient detail for intersectional analysis, and (C3) whether the materials are collected and maintained with human oversight.

To evaluate \emph{preservation of original materials}, we examined whether a source retains the full text of incident reports or instead replaces them with summaries. To evaluate \emph{analytical detail}, we assessed whether the incident data provide enough information to identify who was harmed and how identity-related information is disclosed, either explicitly or implicitly, following the distinction introduced in prior systematic reviews \cite{IntersectionalHCI_2017, Bauer_2021}. To evaluate \emph{human oversight}, we examined whether the source describes editorial review during data submission and validation. These sources performed as follows across our criteria:

\begin{description}[leftmargin=0pt, labelindent=0pt, style=unboxed]
    \item[\textbf{C1: Preservation of original materials.}]  
    AIM represents incidents using LLM-generated summaries rather than preserving full report texts. AIAAIC records incidents primarily as short descriptions written by contributors and accompanied by links to external sources, but does not retain the full text of original materials. In contrast, the AIID preserves the full text of publicly available incident reports by requiring contributors to submit report content and verify that it matches the linked source.
    
    \item[\textbf{C2: Analytical detail.}]  
    AIM aggregates both AI hazards (where harm is only potential) and incidents (where harm has occurred), requiring an additional filtering step to determine whether harm occurred and who was affected. AIAAIC entries vary in detail and are often too brief to support consistent subject-level analysis, depending on how contributors describe incidents. The AIID provides the richest textual detail by preserving full report texts.
    
    \item[\textbf{C3: Human oversight.}]  
    AIM relies largely on automated processes to generate summaries, with limited documented human editorial review. AIAAIC depends on volunteer submissions with minimal standardization beyond link provision. The AIID applies documented editorial review to report submission and incident linking, with automated tools used to detect duplicate reports, and final decisions made through human judgment.
\end{description}

\section{List of Identity Categories, Their Exemplary Values, and Rules for Grouping Identity Values Within Identity Categories}

\label{appendix:categories_list}

\subsection{Listing Identity Categories and Their Exemplary Values}

We constructed this list iteratively by reviewing both identity categories and their exemplary values. We began with race, gender, and class, as identified in earliest intersectionality scholarship \cite{CombaheeRiverCollective1982, Crenshaw1991}, drawing directly on the values used in that work such ``people of color'' and ``white people'', ``women'' and ``men'', and ``working-class'' and ``upper-class''. We then added categories and values introduced in later work, including sexuality and nationality \cite{HillCollins2002}, age, disability, religion, education, language, appearance, and reproductive status \cite{Pauly_Wheel_1996}. To include categories relevant to contemporary AI-related harms, we further expanded the list to include body size, gender identity, immigration status \cite{Goodwill_2021}, indigeneity, caste, political identity, health status, family status, geographic location \cite{Bauer_2021}, and neurodiversity \cite{Neurodiversity_2025}. Finally, to capture incidents affecting non-human entities, we included specie category related to environmental and ecological harm \cite{Hagendorff_2022}. This process resulted in a final list of 26 identity categories paired with exemplary values (see next page).

\newpage
\noindent \textbf{List of Identity Categories and Their Exemplary Values}
\begin{enumerate}[leftmargin=2.1em]
    \item \textbf{Race} (e.g., White, Black)
    \item \textbf{Gender} (e.g., Male, Female)
    \item \textbf{Gender Identity} (e.g., Cisgender, Trans)
    \item \textbf{Class} (e.g., Upper class, Working class)
    \item \textbf{Sexuality} (e.g., Heterosexual, Gay)
    \item \textbf{Nationality} (e.g., German, Syrian)
    \item \textbf{Ability} (e.g., Able-bodied, Disabled)
    \item \textbf{Gender Expression} (e.g., Masculine, Feminine, gender nonconforming)
    \item \textbf{Heritage} (e.g., European descent, African American, Indigenous heritage, diasporic)
    \item \textbf{Age} (e.g., Teenager, Adult, Middle-aged, Senior)
    \item \textbf{Appearance} (e.g., Conventionally attractive, perceived as unattractive)
    \item \textbf{Language} (e.g., Anglophone, English as a second language)
    \item \textbf{Skin Tone} (e.g., Light, Dark)
    \item \textbf{Religion} (e.g., Christian, Muslim)
    \item \textbf{Reproductive Status} (e.g., Fertile, Infertile)
    \item \textbf{Body Size} (e.g., Thin, fat, obese)
    \item \textbf{Education} (e.g., Student, professor, vocational trainee, graduate of an elite university, self-taught)
    \item \textbf{Immigration Status} (e.g., Citizen, permanent resident, temporary visa holder, undocumented migrant)
    \item \textbf{Geography} (e.g., Urban, rural, remote region, informal settlement, university town, capital city)
    \item \textbf{Indigeneity} (e.g., Indigenous person, settler descendant, colonizer lineage)
    \item \textbf{Family Status} (e.g., Single, married, divorced, single parent, caregiver)
    \item \textbf{Caste} (e.g., Brahmin, Dalit)
    \item \textbf{Political Identity} (e.g., Progressive, conservative, libertarian, socialist)
    \item \textbf{Neurodiversity} (e.g., Neurotypical, autistic, ADHD)
    \item \textbf{Health Status} (e.g., Mentally well, living with depression, chronically ill)
    \item \textbf{Species} (e.g., Human, nonhuman animal, plant, insect)
\end{enumerate}

\bigskip
\subsection{Defining Rules for Grouping Identity Values Within Identity Categories}

To ensure consistency and comparability across incidents, we defined rules for grouping identity values within four identity categories: race, gender, class, and age. For race, we grouped values into two categories, distinguishing between ``people of color'' and ``white subjects'', following conventions in prior work on AI harms \cite{FearsHopes2025}. For gender, we grouped values into three categories (female, male, and other), where ``other'' captures non-binary gender identities. For class, we followed the definition by \citet{Ames2011}, which conceptualizes class as ``a category defined by a nexus of income level, educational attainment, and type of employment''. Because education was treated as a separate identity category in our rubric (see point 17 above), we grouped income and occupation values into three categories: lower, middle, and upper class. For example, ``gig worker'' was mapped to lower, ``small business owner'' to middle, and ``politician'' to upper class. For age, we grouped values into five ranges: children (1-9 years), adolescents (10-19 years), younger adults (20-24 years), adults (25-59 years), and older adults (60-99 years).
\clearpage

\section{LLM Prompt Operationalizing the Rubric}
\label{appendix:rubric_prompt}

\vspace{0.65cm}
\tikzstyle{background rectangle}=[thick, draw=black, rounded corners]
\begin{tikzpicture}[show background rectangle]
\node[align=justify, text width=45em, inner sep=1em]{
    \scriptsize 
    
    \noindent\textbf{Persona:} You are an expert AI Incident Analyst. Your core expertise is the application of **Kimberle Crenshaw's intersectionality theory** to analyze AI incident reports. You are precise, context-sensitive, and you avoid flattening identities into isolated categories. To assess how identity contributes to harm, you reason causally and structurally. You often work backwards from the observed harm to trace contributing design choices or detection failures. This approach is similar to Fault Tree Analysis, where analysts start with a failure and identify underlying conditions or assumptions that allowed it to occur. \\
    
    \noindent\textbf{Introduction:} I will provide you with a set of AI incident reports. You will perform four tasks on each report. \\
    
    \noindent\textbf{Tasks:} 
    \\TASK 1: Extract AI Subject Details
    Analyze each report to identify and categorize unique AI Subjects - living entities subjected to or affected by the AI system use. Each AI Subject is identified by identity markers, which are specific attributes (e.g., ``White'' for Race, ``Female'' for Gender) drawn from 26 predefined identity categories rooted in intersectionality theory. To do the task, follow the subtasks 1.1 and 1.2.
    
    If multiple reports refer to the same AI Subject, merge them into a single one by comparing names and identity context. Consider two AI Subjects the same if:
    - Their names refer to the same group (even with different phrasing, e.g., ``Users of Alice'', ``Users of Yandex's Alice'').
    - Their identity markers are identical or overlapping, e.g., two reports list the AI Subjects as ``Young, Spanish-speaking user'' and ``Teenage users who speak Spanish''. Since both share the same markers for Age (``Teenager/Young'') and Language (``Spanish''), they can be treated as the same AI Subject. \\

    Subtask 1.1: Identify the AI Subject
    Extract the exact name of the living entity (human individuals, groups, societies, organizations, or nature) affected by the AI, verbatim as it appears in the report. Exclude inanimate objects (e.g., AI systems, websites, recommender systems, AI agents). \\

    Subtask 1.2: Classify AI Subject Type
    Assign each AI Subject to one category:
    1. An individual - a single person or named entity (e.g., ``John Doe'', ``22-year-old'') 
    2. A group of persons - a subset of people within a specific context (e.g., ``Australian students'', ``employees'', ``protesters'', ``Amazon delivery drivers'') 
    3. Society - a broad, regional, national, or global population (e.g., ``Australians'', ``Canadian public'', ``Moldovan citizens''). If the reference is to a broad national or regional public, it belongs here. If it refers to a specific subgroup, classify under ``a group of persons''. 
    4. Organization - an institution, corporation, company, NGO, government body, university, or media agency (e.g., ``Google'', ``UNICEF'', ``The Guardian'', ``Harvard University'', ``Amazon'') .
    5. Nature - animals, plants, or ecological entities (e.g., ``rhinos'', ``Amazon rainforest'').  
    6. Other - ambiguous or unclear living entities not fitting the above categories. \\
    
    TASK 2: Extract Identity Markers
    For each AI Subject, systematically extract identity markers explicitly or implicitly appearing in the incident across 26 predefined identity categories rooted in intersectionality theory. These markers are specific attributes within a category. Example: ``White'' is a marker within the category of ``Race''.  \\

    \textbf{[List of identity categories and their exemplary values]} as shown in Appendix \ref{appendix:categories_list}\\
    
    Extraction Rules:

    Rule 1: Explicit Markers
    If a marker is explicitly mentioned in the report, return:  
    - Category: The predefined identity category (e.g., ``Gender'') 
    - Marker: The exact marker wording from the report (e.g., ``Non-binary'') 
    - Marker type: ``Explicit''
    - Source: Direct excerpt from the report \\
    Rule 2: Inferred Markers
    If a marker is not explicitly mentioned but can be reasonably inferred from context, return: 
    - Category: The predefined identity category (e.g., ``Family Status'')  
    - Marker: e.g., ``Caregiver''
    - Marker type: ``Inferred''
    - Source: Brief reasoning (1-2 sentences) explaining the inference, citing specific report details. 

    Do not generalize or modify markers; use report-specific evidence only.
    Example: For ``applicants with young children were denied flexible work options`` infer ``Family Status'': ``Caregiver'', with reasoning: ``The phrase 'applicants with young children' indicates that the AI Subject is a caregiver, as it directly references their responsibility for children''. \\
    Rule 3: Non-Mentioned Markers
    If a marker is neither mentioned nor inferable, return: 
    - Category: The predefined identity category (e.g., ``Gender'') 
    - Marker: ``Not mentioned''
    - Marker type: ``None''
    - Source: ``None'' \\

    TASK 3: Assess Causal Relevance of Identity Marker to the Incident
    
    For each identity marker extracted in TASK 2 (explicit or inferred), assess whether the marker was causally relevant to the AI-related harm described in the report.
    To do so, answer the following two binary questions for each extracted marker, and provide a single reasoning explaining these two related questions:
    - Question 1 (Direct Cause): Did this incident happen because the AI Subject was [marker]?
    Return ``Yes'' or ``No''
    Return ``Yes'' only if the marker materially contributed to the system's harmful behavior - that is, the system's decision, action, or failure changed because the subject had this identity marker. Do not return ``Yes'' for general errors or unrelated failures.
    - Question 2 (Alternate Explanation): Would this incident still have happened if the AI Subject was not [marker]? To reason about this one, imagine a version of the AI Subject who is identical in all respects except for this one identity marker (e.g., Gender = Female instead of Male). Would the AI system still have produced the same harmful outcome for this alternative subject?
    Return ``Yes'' or ``No''
    Return ``Yes'' if an alternative subject that is identical in all respects except for this identity marker (e.g., different gender, race, or age) would still have experienced the same harm.
    Return ``No'' if changing the identity marker would likely have prevented or changed the outcome.

    Provide the joint reasoning. Work backwards from the harm to explain how this identity marker may have contributed to it. Focus on the AI system's assumptions, design decisions, or detection failures. To guide your reasoning, consider:
    - Did the system fail to detect, prioritize, or adapt to something about this identity?
    - Was the harm caused by thresholds, classification logic, or other settings that excluded or misrepresented this identity?
    - Did the identity marker influence the wording, tone, or believability of the harmful output?
        
};
\node[xshift=0.5ex, yshift=1ex, overlay, fill=black, text=white, draw=black, rounded corners, right=0.95cm, below=-0.3cm, inner xsep=0.55em, inner ysep=0.32em] at (current bounding box.north west) {
\textit{Part 1/3}
};
\end{tikzpicture}

\vspace{0.65cm}
\tikzstyle{background rectangle}=[thick, draw=black, rounded corners]
\begin{tikzpicture}[show background rectangle]
\node[align=justify, text width=45em, inner sep=1em]{
    \scriptsize
    Examples:
    
    Incident report 1: A security robot at Stanford Mall malfunctioned and knocked over a 16-month-old boy, running over his leg. The robot was designed to detect abnormal noises, environmental changes, and known criminals.
    Extracted Identity Markers: Age = 16-month-old, Gender = Boy

    Q1: Did this incident happen because the AI Subject was 16-month-old? ``Yes''
    Q2: Would this incident still have happened if the AI Subject was not 16-month-old? ``No''\\
    Reasoning: Young children are less detectable by sensors or may behave in unpredictable ways. The robot likely failed to detect or properly respond to a toddler, making the subject's age a causal factor.

    Q1: Did this incident happen because the AI Subject was a boy? ``No''
    Q2: Would this incident still have happened if the AI Subject was not a boy? ``Yes''\\
    Reasoning: The robot did not act differently based on the child's gender. If the child had been a girl, the event would likely have occurred in the same way.\\

    Incident report 2: An AI system capable of generating deepfake videos was deployed in the context of immigration-related communication to impersonate a Toronto-based lawyer. However, it malfunctioned by creating fraudulent videos that targeted newcomers to Canada, exploiting their vulnerability and confusion about immigration rules. 
    Extracted Identity Markers: Immigration Status = Newcomer to Canada, Geography = Toronto-based

    Q1: Did this incident happen because the AI Subject was a newcomer to Canada? ``Yes''
    Q2: Would this incident still have happened if the AI Subject was not a newcomer to Canada? ``No''
    Reasoning: The AI system targeted individuals perceived as unfamiliar with immigration procedures. The harm—deception and fraud—relied on exploiting the uncertainty and vulnerability linked specifically to being a newcomer. If the subject had not been a newcomer (e.g., a long-term resident), the deepfakes would likely not have been directed at them or would have been less effective.

    Q1: Did this incident happen because the AI Subject was Toronto-based? ``No''
    Q2: Would this incident still have happened if the AI Subject was not Toronto-based? ``Yes''
    Reasoning: While the impersonated lawyer was based in Toronto, this detail served primarily to lend superficial credibility. The incident would likely have occurred in the same way if the lawyer had been based in any other Canadian city. The geographic detail did not causally determine the harm experienced by the subjects. \\

    Only produce a MarkerHarm sentence when DirectScore is ``Yes'' and AlternateScore is ``No''.
    In those cases, the MarkerHarm must describe one concrete harmful event from the report that occurred because of that specific identity marker, written in past tense.
    
    For all other combinations of scores, return ``'' (an empty string) for MarkerHarm.\\

    Examples:
    Nationality: ``Ghanaian voters saw false claims about their presidential candidates''.
    Political Identity: ``Opposition supporters saw AI-generated posts falsely accusing their candidate of corruption''.\\

    Here are the AI incident reports: 
    IncidentID: ``{}'', 
    TotalReportNumber: ``{}'', 
    Reports: ``{}'', 

    Return the task results the following JSON format: \\
        
\begin{lstlisting}
        {{"IncidentID": "{{}}", 
        "Description": "[AI system name] was deployed in [context] to [intended function]. However, it [malfunctioned in a way that affected AI Subject]. As a result, [AI Subject] experienced [specific consequences]",
        "ReportNumber": TotalReportNumber, 
        "AI_Subjects": {{ "S1": 
            {{ 
            "SubjectID": "IncidentID" + "-S1", 
            "ReportID": report_number, 
            "Name": "The name of the living entity that is subject to or affected by AI system use", 
            "Type": "Individual" / "Group of persons" / "Society" / "Organizations" / "Nature" / "Other", 
            "Categories": {{ 
                "Race": {{ 
                    "Marker": "Extracted or Inferred race marker", 
                    "MarkerType": "Extracted" / "Inferred" / "None",
                    "Source": "Direct excerpt from the report or brief reasoning explaining the marker inference, citing specific report details",
                    "DirectScore": "Yes" / "No",
                    "AlternateScore": "Yes" / "No",
                    "Reasoning": "If DirectScore is Yes: Briefly explain how the harm traces back to a system behavior or design choice that was sensitive to this identity marker. Use backward reasoning (e.g., detection failure - identity-linked trait - design assumption). Leave empty if DirectScore is No.",
                    "MarkerHarm": "One short sentence naming the exact harmful outcome that actually occurred to subjects with this identity marker in this incident, stated concretely with no abstractions or generalities."
                }}, 
                ..., 

        \end{lstlisting}

};
\node[xshift=0.5ex, yshift=1ex, overlay, fill=black, text=white, draw=black, rounded corners, right=0.95cm, below=-0.3cm, inner xsep=0.55em, inner ysep=0.32em] at (current bounding box.north west) {
\textit{Part 2/3}
};
\end{tikzpicture}

\vspace{0.65cm}
\tikzstyle{background rectangle}=[thick, draw=black, rounded corners]
\begin{tikzpicture}[show background rectangle]
\node[align=justify, text width=45em, inner sep=1em]{
    \scriptsize 

        \begin{lstlisting}
                ..., 
                "Specie": {{ 
                    "Marker": "Extracted or Inferred specie marker", 
                    "MarkerType": "Extracted" / "Inferred" / "None",
                    "Source": "Direct excerpt from the report or brief reasoning explaining the marker inference, citing specific report details",
                    "DirectScore": "Yes" / "No",
                    "AlternateScore": "Yes" / "No",
                    "Reasoning": "If DirectScore is Yes: Briefly explain how the harm traces back to a system behavior or design choice that was sensitive to this identity marker. Use backward reasoning (e.g., detection failure - identity-linked trait - design assumption). Leave empty if DirectScore is No.",
                    "MarkerHarm": "One short sentence naming the exact harmful outcome that actually occurred to subjects with this identity marker in this incident, stated concretely with no abstractions or generalities."
                }} 
            }},
            ..., "SN": {{ 
            "SubjectID": "IncidentID" + "-SN", 
            "ReportID": report_number, 
            "Name": "The name of the living entity that is subject to or affected by AI system use", 
            "Type": "Individual" / "Group of persons" / "Society" / "Organizations" / "Nature" / "Other", 
            "Categories": {{ 
                "Race": {{ 
                    "Marker": "Extracted or Inferred race marker", 
                    "MarkerType": "Extracted" / "Inferred" / "None",
                    "Source": "Direct excerpt from the report or brief reasoning explaining the marker inference, citing specific report details",
                    "DirectScore": "Yes" / "No",
                    "AlternateScore": "Yes" / "No",
                    "Reasoning": "If DirectScore is Yes: Briefly explain how the harm traces back to a system behavior or design choice that was sensitive to this identity marker. Use backward reasoning (e.g., detection failure - identity-linked trait - design assumption). Leave empty if DirectScore is No.",
                    "MarkerHarm": "One short sentence naming the exact harmful outcome that actually occurred to subjects with this identity marker in this incident, stated concretely with no abstractions or generalities."
                }}, ..., 
                "Specie": {{ 
                    "Marker": "Extracted or Inferred specie marker", 
                    "MarkerType": "Extracted" / "Inferred" / "None",
                    "Source": "Direct excerpt from the report or brief reasoning explaining the marker inference, citing specific report details",
                    "DirectScore": "Yes" / "No",
                    "AlternateScore": "Yes" / "No",
                    "Reasoning": "If DirectScore is Yes: Briefly explain how the harm traces back to a system behavior or design choice that was sensitive to this identity marker. Use backward reasoning (e.g., detection failure - identity-linked trait - design assumption). Leave empty if DirectScore is No.",
                    "MarkerHarm": "One short sentence naming the exact harmful outcome that actually occurred to subjects with this identity marker in this incident, stated concretely with no abstractions or generalities."
                }} }}, }} }} }} 
        \end{lstlisting}

        JSON Formatting Rules: \\
        Rule 1: Go systematically through each of the ***reports***. 
        Rule 2: Go systematically through each of the ***26 predefined identity categories***. 
        Rule 3: Separate each identity category with a comma, including the last category. 
        Rule 4: Ensure that the JSON response strictly follows the format with proper commas between AI subjects and identity categories. 
        Rule 5: Before final output, deduplicate AI Subjects across all reports in the incident based on their Name and identity attributes. Return only one entry per unique subject, even if it appears in multiple reports.
        Rule 6: Output only JSON - do not include any other content. 

};
\node[xshift=0.5ex, yshift=1ex, overlay, fill=black, text=white, draw=black, rounded corners, right=0.95cm, below=-0.3cm, inner xsep=0.55em, inner ysep=0.32em] at (current bounding box.north west) {
\textit{Part 3/3}
};
\end{tikzpicture}

\clearpage

\section{Validation of the Rubric Results through LLM Misattributions}
\label{appendix:misattribution}

\begin{figure}[h!]
    \centering
    \includegraphics[width=\linewidth]{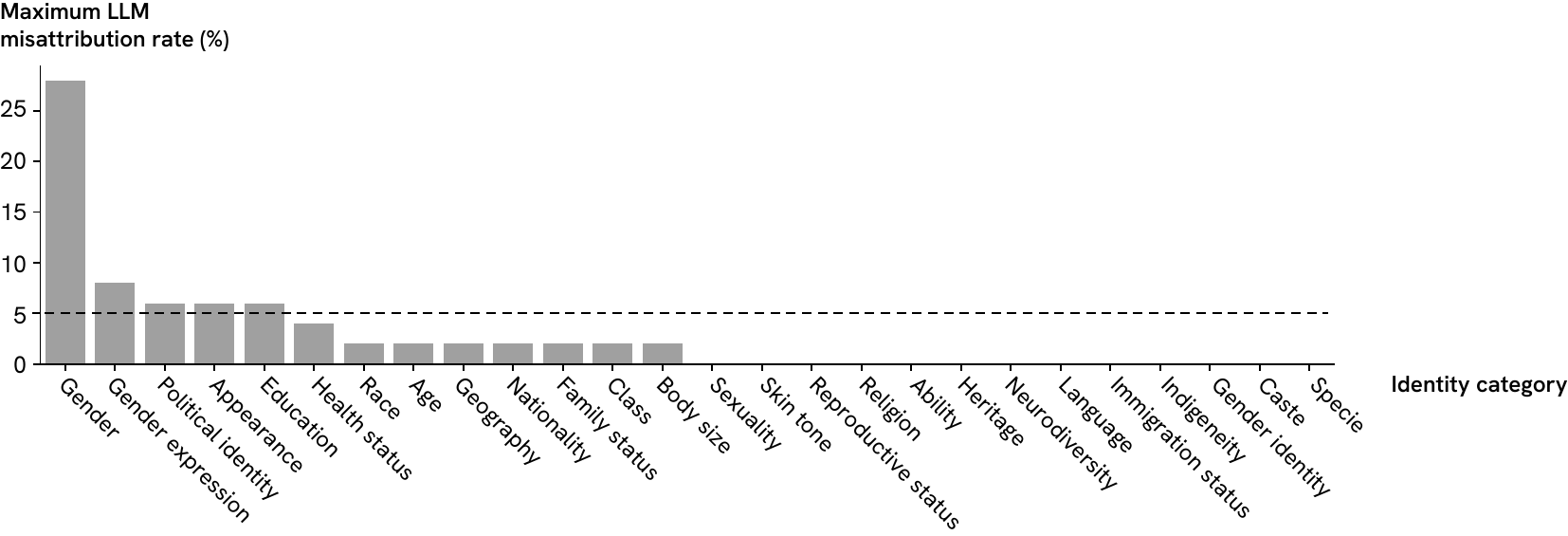}
    \caption{\textbf{Maximum misattribution rate across annotators and the LLM by identity category.} For each category, we report the highest error rate observed across two annotators and the LLM, providing a conservative estimate of misattributions --- cases where the LLM's identity category assignments differed from those of the annotators. Misattributions were generally low across categories (0--5\%), with the exception of gender (28\%). The majority of gender misattribution cases (90\%) arose when annotators inferred gender from implicit cues (e.g., pronouns, gender-coded titles, or names), whereas the LLM behaved conservatively and did not infer it.}
    \Description{A bar chart showing the maximum LLM misattribution rate (as a percentage) for 25 identity categories, sorted in descending order. The y-axis ranges from 0 to approximately 28 percent. A horizontal dashed line marks the 5 percent threshold.
    Gender has by far the highest misattribution rate at approximately 28 percent, well above the dashed line. Gender expression is the second highest at approximately 8 percent, also above the line. Political identity and Appearance are both at approximately 6 percent, just above the line. Education is at approximately 4 percent, just below the line. Health status, Race, Age, Geography, Nationality, Family status, Class, and Body size all fall between approximately 1 and 2 percent. The remaining categories --- Sexuality, Skin tone, Reproductive status, Religion, Ability, Heritage, Neurodiversity, Language, Immigration status, Indigeneity, Gender identity, Caste, and Species --- have misattribution rates at or near 0 percent, with no visible bars.}
    \label{fig:misattribution}
\end{figure}

\section{Example AI Incidents Illustrating Intersections Between Identity Categories}
\label{appendix:incident_table}

\begingroup

\footnotesize
\renewcommand{\arraystretch}{1.35}
\begin{longtable}
{|p{2.8cm}|c|p{7.5cm}|p{3.6cm}|}

\caption{\textbf{Example AI incidents illustrating the most prevalent intersections between identity categories.} Each row corresponds to one cell of the heatmap in Figure~\ref{fig:heatmap}.}
\label{tab:intersections} \\
\hline
\textbf{Intersection} & \textbf{ID} & \textbf{Incident description and source} & \textbf{Harmed subjects} \\
\hline
\endfirsthead
\multicolumn{4}{c}{\tablename\ \thetable{} -- continued} \\
\hline
\textbf{Intersection} & \textbf{ID} & \textbf{Description} & \textbf{Subjects} \\
\hline
\endhead
\hline \endfoot
\hline \endlastfoot
\textbf{Age + Gender}\newline
(73 incidents with 125
\newline intersectional subjects) & 188 & The Technology Platform for Social Intervention was deployed in the Argentine province of Salta to predict which specific low‑income girls would become pregnant as adolescents. However, it targeted and surveilled marginalized girls and young women using invasive demographic and socioeconomic data, labeling them as ``predestined'' for teen pregnancy. As a result, these girls and women experienced violations of privacy, stigmatization, and coercive surveillance tied to access to essential social services \cite{AIID_Incident188_2018}. & Age: Adolescents,\newline
Gender: Female (girls and women from low-income areas in the province of Salta, including women and girls between the ages of 10 and 19, many from migrant families and Indigenous Wichí, Qulla, and Guaraní communities) \\
\hline
\textbf{Age + Class}\newline
(24 incidents with 32\newline
intersectional subjects) & 1144 & Elon Musk's xAI supercomputer facility ``Colossus'' was deployed in South Memphis to provide compute power for the Grok AI chatbot. However, it was powered with large numbers of methane gas turbines. Many of them operated without permits, in a way that concentrated toxic air pollution and heavy resource use around nearby residential neighborhoods. For example, ozone levels exceeded the Environmental Protection Agency's safety standards and sensitive groups, like children and adults with respiratory issues, were advised not to go outside. As a result, residents of poor, historically Black South Memphis neighborhoods experienced increased health and environmental risks while being excluded from decision-making. \cite{AIID_Incident1144} & Age: Children,\newline
Class: Lower (South Memphis residents) \\
\hline
\textbf{Age + Nationality}\newline 
(27 incidents with 40\newline 
intersectional subjects) & 101 & Fraud‑detection and welfare risk‑scoring algorithms were deployed in Dutch tax and welfare administration to detect or predict social and childcare benefits fraud. However, they systematically labeled certain beneficiaries as high‑risk or fraudulent, triggering punitive debt collection and investigations that disproportionately targeted people with dual nationalities, ethnic minorities, immigrants, low‑income families, and other vulnerable recipients. As a result, these subjects were wrongly accused of fraud, pushed into poverty and debt, lost benefits, in some cases lost their children, and suffered severe psychological and social harms \cite{AIID_Incident101}. & Age: Adults,\newline
Nationality: Iraqi \\
\hline
\textbf{Age + Political Identity}\newline (20 incidents with 27\newline intersectional subjects) & 202 & AI deepfake avatars of South Korean presidential candidates were deployed in election campaigns to appeal to younger voters and extend candidates' reach. However, they potentially misled parts of the electorate by masking candidates' real traits, using deceptive framing, and risking misuse for fake political content, affecting voters' ability to evaluate genuine candidates and information. As a result, young South Korean voters and the broader Korean public experienced an increased risk of political deception and erosion of trust in media and democratic processes \cite{AIID_Incident202}. & Age: Adults, Young adults,\newline Political Identity: Voters, especially swing voters in their 20s and 30s \\
\hline
\textbf{Age + Race}\newline (36 incidents with 47 \newline intersectional subjects) & 40 & COMPAS and similar recidivism risk algorithms were deployed in criminal justice contexts to predict defendants' likelihood of reoffending and inform bail, sentencing, parole, and supervision decisions. However, they produced racially disparate error patterns, opaque and potentially inaccurate scores, and trade‑secret‑protected outputs that defendants and judges could not effectively scrutinize or contest. As a result, criminal defendants and incarcerated or paroled people experienced harsher or otherwise unjust decisions, including higher risk classifications, longer or more restrictive sentences, and parole denials driven by flawed or unchallengeable algorithmic assessments \cite{AIID_Incident40}. & Age: Young adults,\newline Race: Black \\
\hline
\textbf{Class + Gender}\newline (22 incidents with 26\newline intersectional subjects) & 924 & AI-based deepfake generation and targeted ad systems were deployed on social media platforms to drive traffic to scam trading websites and impersonation scams. However, they generated and promoted non-consensual sexually explicit fake images and fabricated articles using celebrities' likenesses in ways that sexualised and misled people. As a result, the celebrities experienced reputational and privacy harms, while social media users were deceived and in some cases lost money to scams \cite{AIID_Incident924}. & Class: Upper (celebrity),\newline Gender: Female (high-profile women including Naga Munchetty, Taylor Swift and Megan Thee Stallion) \\
\hline
\textbf{Class + Nationality}\newline (44 incidents with 81\newline intersectional subjects) & 9 & New York State and New York City value-added teacher evaluation systems were deployed in public schools to measure and rate teacher effectiveness using statistical models on student test scores. However, they produced volatile, inaccurate, and sometimes contradictory ratings, and these scores were publicly released and tied to high-stakes consequences in ways that affected teachers. As a result, individual teachers and groups of teachers experienced unfair low or inconsistent evaluations, public shaming, and potential employment consequences based on flawed data \cite{AIID_Incident9}. & Class: Middle (Academic),\newline Nationality: American \\
\hline
\textbf{Class + Political Identity}\newline (38 incidents with 59\newline intersectional subjects)  & 1077 & AI-generated voice and text content was deployed in impersonation and phishing campaigns to obtain access to personal and official accounts. However, it impersonated senior U.S. officials and contacted current or former senior government officials and their contacts in deceptive ways. As a result, these officials, their contacts, and voters experienced targeted fraud attempts, privacy and security risks, and voter suppression \cite{AIID_Incident1077}. & Class: Upper, \newline Political Identity: Political elite (White House Chief of Staff; senators, governors, and business executives; current or former senior U.S. federal or state government officials and their contacts) \\
\hline
\textbf{Class + Race}\newline (40 incidents with 63\newline intersectional subjects) & 335 & A visa application streaming algorithm was deployed in the UK immigration context to triage entry visa applications using a traffic-light risk system. However, it explicitly incorporated nationality and produced racially biased streams that favored people from rich white countries and disadvantaged applicants from `suspect' nationalities, especially poorer people of color. As a result, many visa applicants from these groups experienced intensive scrutiny, delays, higher refusal rates, and were sometimes unable to visit family, work, study, or attend events in the UK \cite{AIID_Incident335}. & Class: Lower,\newline Race: People of color \\
\hline
\hline
\textbf{Gender + Nationality}\newline (14 incidents with 20\newline intersectional subjects) & 672 & The Lavender AI system (used with tools like Where's Daddy? and The Gospel) was deployed in the context of Israel's war in Gaza to algorithmically identify alleged Hamas and PIJ operatives and their locations for airstrikes. However, it generated mass ``kill lists'' of Palestinian men and linked them to family homes with minimal human review, in a framework that pre-authorized high civilian ``collateral'' deaths. As a result, Palestinian residents of Gaza, particularly men marked as suspects and their families (often women and children), experienced large-scale bombardment of homes, killings, and injury \cite{AIID_Incident672}. & Gender: Male,\newline Nationality: Palestinian \\
\hline
\textbf{Gender + Political Identity} (23 incidents with 37\newline intersectional subjects) & 904 & AI-powered deepfake and ``nudifying'' tools were used on social media and porn platforms to generate non-consensual sexualized images and videos of political activists such as Kate Isaacs. The fabricated content was circulated without consent and often accompanied by threats and harassment. As a result, groups of women experienced image-based sexual abuse, fear for their safety, and ongoing psychological distress \cite{AIID_Incident904}. & Gender: Female,\newline Political Identity: Activist \\
\hline
\textbf{Gender + Race}\newline
(39 incidents with 64\newline intersectional subjects) & 13 & Perspective API, developed by Jigsaw/Google, was deployed in online commenting and content-moderation contexts to score and filter ``toxic'' or ``uncivil'' language. However, it systematically over-scored self-identifications and neutral references to many marginalized or politicized identities as toxic while under-scoring politely phrased bigotry, and failed to capture nuanced uncivility, leading to disproportionate flagging, suppression, or mischaracterization of speech by these groups and skewed measurements of public discourse \cite{AIID_Incident13}. & Gender: Female,\newline Race: Black, Latinx, White (people described with terms like ``gay'', ``genderqueer'', ``deaf'' in comments or news text) \\
\hline
\hline
\textbf{Nationality + Political Identity}\newline (83 incidents with 167\newline intersectional subjects) & 972 & AI-assisted disinformation tools were deployed in the context of the 2024 US election to influence public opinion about Kamala Harris, Tim Walz, and other political figures. However, they generated and amplified fake campaign websites and deepfake videos that misrepresented these individuals and misled voters. As a result, targeted politicians and segments of the US electorate experienced reputational harm and exposure to deceptive content intended to distort democratic decision-making \cite{AIID_Incident972}. & Nationality: American,\newline Political Identity: Political elite (Vice President Kamala Harris; Governor Tim Walz; Representative Barry Moore, Senator Marco Rubio, Senator Marsha Blackburn, and Representative Michael McCaul) \\
\hline
\textbf{Nationality + Race}\newline (29 incidents with 47\newline intersectional subjects) & 650 & AI text-to-image systems were used in a pre-election disinformation context to generate deepfake photographs depicting Donald Trump with Black voters and civil rights figures in order to influence political attitudes. However, these AI-generated images misrepresented Black voters' political support and manipulated perceptions of Black political behavior. As a result, Black voters and the broader public experienced deceptive visual propaganda about Black political alignment \cite{AIID_Incident650}. & Nationality: American,\newline Race: Black \\
\hline
\hline
\textbf{Political Identity + Race}\newline (25 incidents with 32\newline intersectional subjects)  & 1075 & AI-powered facial recognition was deployed in New Orleans via Project NOLA's live camera network to continuously scan public streets for suspects and trigger real-time alerts to police. However, it was operated in secret, outside mandated oversight and reporting requirements, in a way that affected residents, visitors, and people on or added to watchlists. As a result, these subjects experienced continuous warrantless surveillance, risk of misidentification, and arrests and detentions without transparency or due process protections \cite{AIID_Incident1075}. & Political Identity: Activists (and others speaking out or challenging government policies),\newline Race: People of color \\
\end{longtable}

\endgroup

\end{document}